\documentclass{svjour3}                     
\smartqed  
\journalname{Experimental Astronomy}
\usepackage{mathptmx}      
\usepackage{graphicx}
\usepackage[latin1]{inputenc}
\usepackage[square,numbers]{natbib}

\newcommand{\arcsec}{$^{\prime\prime}$}
\newcounter{IonCS}
\newcommand{\ion}[2]{\setcounter{IonCS}{#2}#1\,{\small{\Roman{IonCS}}}}
\newcommand{\sect}[1]{Sect.\,\ref{#1}}

\newcommand{\fig}[1]{Fig.\,\ref{#1}}

\newcommand{\tab}[1]{Table\,\ref{#1}}
\newcommand{\Lya}{Ly-$\alpha$}
\begin{document}

\title{Magnetic Imaging of the Outer Solar Atmosphere (MImOSA)
}

\subtitle{Unlocking the driver of the dynamics in the upper solar atmosphere}


\author{\mbox{H. Peter}
\and	\mbox{E. Alsina Ballester}
\and	\mbox{V. Andretta}
\and	\mbox{F. Auch\`ere}
\and	\mbox{L. Belluzzi}
\and	\mbox{A. Bemporad}
\and	\mbox{D. Berghmans}
\and	\mbox{E. Buchlin}
\and	\mbox{A. Calcines}
\and	\mbox{L.P. Chitta}
\and	\mbox{K. Dalmasse}
\and	\mbox{T. del Pino Alem\'an}
\and	\mbox{A. Feller}
\and	\mbox{C. Froment}
\and	\mbox{R. Harrison}
\and	\mbox{M. Janvier}
\and	\mbox{S. Matthews}
\and	\mbox{S. Parenti}
\and	\mbox{D. Przybylski}
\and	\mbox{S.K. Solanki}
\and	\mbox{J. \v{S}t\v{e}p\'an}
\and	\mbox{L. Teriaca}
\and	\mbox{J. Trujillo Bueno}
}


\institute{H. Peter, L.P. Chitta, A. Feller, D. Przybylski, L. Teriaca
\at		Max Planck Institute for Solar System Research, G\"ottingen, Germany
\and	E. Alsina Ballester
\at		Istituto Ricerche Solari Locarno, Locarno-Monti, Switzerland
\and	L. Belluzzi
\at		Istituto Ricerche Solari Locarno, Locarno-Monti, Switzerland
        and Leibniz-Institut f{\"u}r Sonnenphysik, Freiburg, Germany
\and	V. Andretta 
\at		INAF Osservatorio Astronomico di Capodimonte, Napoli, Italy 
\and	F. Auch\`ere, E. Buchlin, M. Janvier
\at		Universit\'e Paris-Saclay, CNRS, Institut d'Astrophysique Spatiale, 91405 Orsay, France
\and	A. Bemporad
\at		INAF Osservatorio Astrofisico di Torino, Italy
\and	D. Berghmans
\at		Royal Observatory of Belgium, Brussels, Belgium
\and	A. Calcines
\at		Centre for Advanced Instrumentation, Durham University, UK	
\and	K. Dalmasse
\at		IRAP, Universit\'e de Toulouse, CNRS, CNES, UPS, Toulouse, France
\and	T. del Pino Alem\'an, J. Trujillo-Bueno
\at		Instituto de Astrof\'isica de Canarias, Tenerife, Spain
\and	C. Froment
\at		LPC2E, CNRS/University of Orléans/CNES, Orléans, France
        and Institute of Theoretical Astrophysics, University of Oslo, Norway
\and	R. Harrison
\at		RAL Space, STFC Rutherford Appleton Laboratory, Oxfordshire, UK
\and	S. Matthews
\at		UCL Mullard Space Science Laboratory, Surrey, UK
\and	S. Parenti
\at		Institut d'Astrophysique Spatiale, Orsay, France
        and D\'epartement d'Astrophysique/AIM, CEA/IRFU, CNRS/INSU, Universit\'e Paris-Saclay, Universit\'e de Paris, 91191 Gif-sur-Yvette, France
\and    S.K. Solanki
\at		Max Planck Institute for Solar System Research, G\"ottingen, Germany
        and
        School of Space Research, Kyung Hee University, Yongin, Republic of Korea
\and	J. \v{S}t\v{e}p\'an
\at		Astronomical Institute ASCR, Ond\v{r}ejov, Czech Republic
}

\date{draft version: \today}

\maketitle

\begin{abstract}
The magnetic activity of the Sun directly impacts the Earth and human life. Likewise, other stars will have an impact on the habitability of planets orbiting these host stars. Although the magnetic field at the surface of the Sun is reasonably well characterised by observations, the information on the magnetic field in the higher atmospheric layers is mainly indirect. This lack of information hampers our progress in understanding solar magnetic activity. Overcoming this limitation would allow us to address four paramount long-standing questions:
(1) How does the magnetic field couple the different layers of
the atmosphere, and how does it transport energy?
(2) How does the magnetic field structure, drive and interact with the plasma in the chromosphere and upper
atmosphere?
(3) How does the magnetic field destabilise the outer solar atmosphere
and thus affect the interplanetary environment?
(4) How do magnetic processes accelerate particles to high energies?
New ground-breaking observations are needed to address these science questions. We suggest a suite of three instruments that far exceed current capabilities in terms of spatial resolution, light-gathering power, and polarimetric performance:
(a) A large-aperture UV-to-IR telescope of the 1-3 m class aimed mainly to measure the magnetic field in the chromosphere by combining high spatial resolution and high sensitivity.
(b) An extreme-UV-to-IR coronagraph that is designed to measure the large-scale magnetic field in the corona with an aperture of about 40\,cm.
(c) An extreme-UV imaging polarimeter based on a 30\,cm telescope that combines high throughput in the extreme UV with polarimetry to connect the magnetic measurements of the other two instruments.
Placed in a near-Earth orbit, the data downlink would be maximised, while a location at L4 or L5 would provide stereoscopic observations of the Sun in combination with Earth-based observatories. This mission to measure the magnetic field will finally unlock the driver of the dynamics in the outer solar atmosphere and thereby will greatly advance our understanding of the Sun and the heliosphere.
\end{abstract}

\section{Introduction}

The magnetic field plays a central role in many if not most of the processes
we observe in our universe.  It governs complex plasma processes from the formation of structures in the early
universe to the interplanetary space that connects
planets to their host stars.
Although we know the magnetic field at the solar surface reasonably well, our information on it in the higher layers is mainly indirect. This state of knowledge is highly unsatisfactory because the field carries most of the energy in these layers and is the dominant driver of most of the phenomena that are observed there (e.g. heating of the gas to millions of degrees, coronal shape and brightness, magnetic reconnection and its effects, and large-scale ejection of parts of the solar atmosphere). The lack of direct measurements of coronal magnetic fields is similar to being blind to the main driver of solar activity. We contend that this is one of the major hindrances to our understanding of coronal physics. Obtaining these measurements would help to explain a very wide variety of phenomena.

We can follow the interaction of the magnetic field and plasma on our central star in detail and 
may essentially use the Sun as a laboratory for uncovering key magnetic and plasma processes that act throughout the universe.
In the coming decades, we expect that answering the following overarching questions will lead to a giant leap in understanding the Sun:

\begin{itemize}

\item[1.] How does the magnetic field couple the different layers of the atmosphere, and how does it transport energy?

\item[2.] How does the magnetic field structure, drive and interact with the plasma
in the chromosphere and upper
atmosphere?

\item[3.] How does the magnetic field destabilise the outer solar atmosphere
and thus affect the interplanetary environment?

\item[4.] How do magnetic processes accelerate particles to high energies?

\end{itemize}

Answers to these questions will be relevant not only to solar research alone,
but will affect a broad range of current and upcoming topics in astrophysics.
The Sun is a template for other stars. This makes it a key to understanding the
host stars of other planets and how these host stars might affect the formation and evolution
of life on their planets. For the reliable detection of Earth-sized planets around other
stars and in particular for their habitability, we have to understand stellar variability.
The magnetic field is also responsible for the UV emission from planet-hosting stars, which is central for the chemistry in the planetary atmosphere, for instance.
If the magnetic field destabilises parts of the stellar atmosphere, which results in the ejection of plasma and magnetic field into interplanetary space, this can also strongly affect planets around the host star.
Finally, high-energy particles generated by magnetic and plasma processes
in the stellar atmosphere can have severe impacts on life. In the case of the the solar system, this includes space technology.

The Solar and Heliospheric Observatory (SOHO) was launched in 1995
\cite[][]{1995SoPh..162....1D}.
The intensive research in the two decades since then has enabled us now to define the questions we outlined above with high precision. Observational
work together with progress in theoretical understanding and numerical modelling
shows clear paths to solving these questions in the coming decades if the observational tools can be made adequately available.

In this paper we describe the key measurements that give direct access to the magnetic fields in the solar chromosphere and upper
atmosphere. These measurements will have to exceed what is currently available. This will enable us to address the overarching questions when we combine it   with the (simultaneous) measurements of the dynamic response of the plasma to changes that are induced by the evolving magnetic field.

\begin{figure}[t]
{\includegraphics[width=\textwidth]{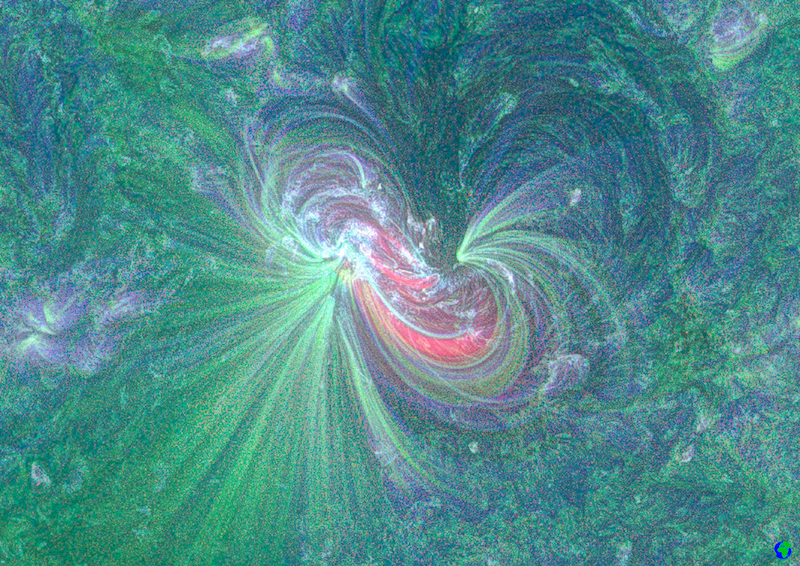}}
\caption{%
Hot plasma loops on the Sun outlining the magnetic field in an active region.
This multi-color image is a composite of AIA 171\,\AA\ (ca.\ 1\,MK; green), AIA 193\,\AA\ (ca.\ 1.5\,MK; blue) and AIA 94\,\AA\ (ca.\ 7\,MK; red).
An image of Earth to scale is overlaid at the bottom right corner.
The image is processed using a Multi-Scale Gaussian Normalization technique \cite[][]{2014SoPh..289.2945M}. 
Data courtesy of NASA/SDO and  AIA science team.
\label{F:cover}
}
\end{figure}

\section{Overarching science questions\label{S:science}}

The focus of this paper is on measuring the magnetic field in the
outer solar atmosphere, i.e. in the upper chromosphere, transition region, and corona. This information is currently not directly available on a routine basis. 
Extrapolations together with extreme UV imaging can give an instructive overview of the large-scale structure of the magnetic field in the outer solar atmosphere, where the relevant physical processes occur (see \fig{F:cover}).
Our understanding of the solar atmosphere is drastically limited by such indirect estimations of the magnetic field, however. 
We have to rely on the theory of extrapolations of the magnetic field that is measured in the photosphere, i.e.\ the visible
surface of the Sun.
These extrapolations have to apply several assumptions, e.g.\ that the solar
atmosphere is a vacuum or that the Lorentz force is the only relevant force
and that the inertia of the plasma is completely negligible. Observations of
the dynamic solar atmosphere clearly show that assumptions like this are not appropriate.
 The actual measurement of the changing magnetic configuration
is a key part in solving the puzzle of how the magnetic field interacts with
the plasma, in particular when fast-changing processes are followed that are driven by
magnetic
reconnection, for instance.

With direct measurements of the magnetic field throughout the solar atmosphere, we will be able to locate sites of transient energy release (through magnetic reconnection), analyse the formation and disruption of current sheets where plasma is heated, study the transport of energy through different modes of magnetohydrodynamic waves (or through energetic particles), and investigate the helicity of erupting structures, all by unveiling how the magnetic field connects the different parts of the atmosphere. These measurements together will provide us with an understanding of solar magnetic activity: how does the magnetic field govern the energetic radiation from the Sun and how does it drive space weather effects? These two questions have implications for Earth, and likewise for planets of other host stars.

The polarisation of spectral lines encodes the information on the magnetic field of the outer solar atmosphere (i.e. the chromosphere and beyond). This encoded information has not been accessible to current space instruments. Sub-orbital rocket flights of about a few minutes each have provided the only glimpse into this domain. 
Because the level of polarisation is generally low, the required noise level has to be as low as a few times 10$^{-4}$ in terms of the intensity for some polarimetric measurements. In order to achieve this signal-to-noise ratio at high spatial resolution, the required instrumentation has to combine large aperture and high throughput. This is as yet unavailable in space.

In the following we describe how the direct measurement of the magnetic
field in the outer solar atmosphere (from the upper chromosphere into the corona) will provide the crucial information
we need to answer the general four questions we listed above. We first concentrate
on the implications for solar research and then address how this will contribute
to a better understanding in other fields of astrophysics.

\subsection{Interaction of magnetic field and plasma}

The interaction of magnetic field and plasma includes a diversity of processes that are currently poorly understood. Of particular interest are the role of the magnetic field in the coupling of the different layers of solar and stellar atmospheres, and the conversion and transport of energy. The most complex mixture of physics on the Sun is found in the chromosphere. In a simple one-dimensional picture, this is the layer above the surface, where the temperature starts to rise outwards. This layer is sandwiched between the photosphere, where most photons originate, and the corona, where the temperature reaches several million Kelvin.

The chromosphere is a highly complex region. The magnetic field energy there is comparable to the internal energy of the gas. This causes an intense turbulent interaction between the two, in particular when the velocities reach the sound speed. To add to the complexity, the radiation processes and line formation are far away from local thermodynamic equilibrium (LTE), and the ionisation is only partial. A model that describes this requires an extended Ohm's law and the consideration of multi-fluid effects.

The hot part of the outer atmosphere, the corona, is magnetically rooted in the chromosphere. To answer fundamental questions on the dynamics and energisation of the corona, we have to understand how the magnetic field couples the corona to the chromosphere. This is essential for describing the internal magnetic state of coronal loops. They are the basic building blocks of the corona. Understanding the magnetic structure of coronal features on the Sun is a stepping stone to the unsolved problem of the heating of hot plasma that is observed throughout the universe.

The large magnetic Reynolds numbers in the chromosphere and corona mean that we can expect magnetic turbulence. The chromosphere and corona are additionally filled with different wave modes that range from acoustic waves to Alfv\'en waves, including all mixed forms. To unravel the magnetic effects on the plasma, we have to decompose these wave modes into their respective parts and relate them to the (slower) build-up of currents that are then dissipated to energise the plasma. 
The key for this is a continuous knowledge of the magnetic field throughout the atmosphere. Only then can we understand the coupling and the transport of energy by the magnetic field.

\begin{figure}[t]
{\includegraphics[width=\textwidth]{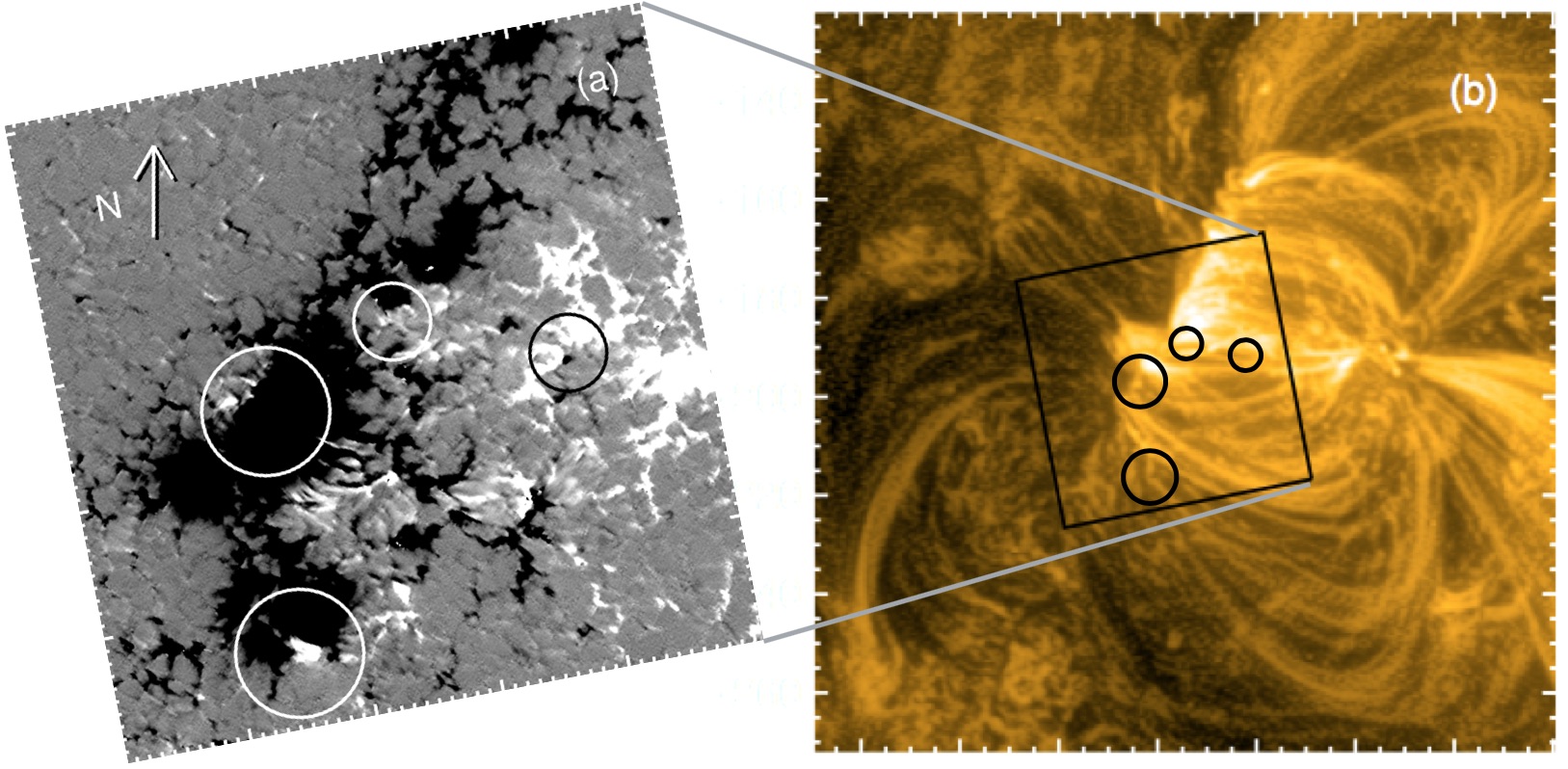}}
\caption{%
Coupling from the photosphere to the corona. The right panel (b) shows 10$^{\,6}$\,K hot plasma loops in the corona in a small active region (150\arcsec$\times$150\arcsec) seen by AIA/SDO. The left panel (a) shows the magnetic field on the surface in the center part of the active region (50\arcsec$\times$50\arcsec). Black and white are opposite magnetic polarities. Small regions with opposite magnetic polarities are found at the footpoints of coronal loops, as marked by circles.
Based on \cite{2017ApJS..229....4C}.
\label{F:coupling}
}
\end{figure}

New observations will have to provide reliable information on the magnetic structure of the upper atmosphere that is not available today and will not be available in the coming years with current instrumentation. We need direct access to the magnetic field through remote-sensing observations to detect the magnetic current sheets that cover a range of spatial and temporal scales \cite[][]{2003Natur.425..692S}. In coronal loops as in chromospheric features, we see fine structure down to the resolution limit of current instruments. Imaging and spectroscopic observations indicate thin structures at scales down to 50\,km \cite[][]{2017ApJS..229....1S}.

The  requirements on spatial resolution for magnetic structure measurements are slightly more relaxed: the magnetic field starts to dominate the plasma in terms of (internal) energy in the upper chromosphere (and beyond). Thus the spatial variations of the magnetic field can be expected to be smaller than those of the plasma in responding to the magnetic driving. To achieve magnetic field measurements, the light needs to be decomposed into its polarisation states. This increases the requirements on the noise level, i.e.\ the collected number of photons. This means that a large aperture is also required for the magnetic measurements.

\subsection{Magnetic coupling in the atmosphere}

To understand the magnetic activity of the Sun, we have to follow the magnetic connections that channel the energy transport throughout the atmosphere. This is a multi-scale problem because large-scale magnetic features, such as long coronal loops or helmet streamers, are rooted in small magnetic patches in the low atmosphere. Recent observations demonstrated the key role of small-scale magnetic patches in the photosphere for large coronal structures \cite[][]{2017ApJS..229....4C}, see \fig{F:coupling}. Even compact flaring loops show small-scale flux cancellation at their foot points \cite[][]{2018A&A...615L...9C}. Only by tracing and quantifying the energy flux deep in the chromosphere can we build a physical connection between the small-scale processes in the photosphere and the large-scale coronal features. The prime requirement to achieve this is to measure the magnetic field in all layers of the solar atmosphere in order to gain insight into the magnetic coupling.

Here we describe with one example how we investigate the magnetic coupling from the chromosphere to the large-scale corona. High-resolution observations enable us to characterise the magnetic field close to the limb. Very stable observing conditions such as are found on a space-based platform are crucial for observations like this. For example, the modest 50\,cm telescope on Hinode was able measure the magnetic field up to heliocentric angles of about 80$^\circ$, and the polar magnetic field could be characterised based on this \cite[][]{2008ApJ...688.1374T}. With higher resolution and sufficient sensitivity, an investigation even closer to the limb would be possible and the magnetic field in the chromosphere could be measured.
This could then be directly connected to coronagraphic observations above the limb. When observations in the extreme UV are added to this, structures can be followed continuously from the chromosphere into the corona, i.e.\ from small to large scales. With this full magnetic field information, magnetic disturbances such as Alfv\'enic perturbations could be traced, and the energy flow from the chromosphere into the corona could be followed.

Another prime target for the study of magnetic connectivity is the destabilisation of the upper solar atmosphere (discussed in more detail in the next subsection). Magnetic structures in the corona can become unstable in a very short time (compared to their overall lifetime) and then erupt. The actual triggers of this process are still elusive. Only a full characterisation of the connectivity of the magnetic field before the onset of the destabilisation can provide the necessary information for understanding the magnetic configuration that later erupts. Here again the coupling of the magnetic field across scales plays a significant role: at least in some scenarios, small-scale processes cut the tethers of the magnetic field that holds the magnetic structure down that later erupts \cite[][]{2001ApJ...552..833M}.

\subsection{Destabilising the outer solar atmosphere}

During large disruptions of the upper solar atmosphere, huge amounts of magnetic energy %
are converted into thermal energy and kinetic energy, and accelerate energetic particles. The associated heating and acceleration processes through magnetic reconnection are still poorly understood, and even the change in magnetic structures in the course of an eruption is not well constrained. Extrapolations of the magnetic field from the surface  are currently used to constrain the change in free energy of the magnetic field. Before and during dynamic events, however, the assumptions that underlie these extrapolations break down. It is therefore mandatory to actually measure the coronal magnetic field before, during, and after the disruption to fully characterise this process.
This destabilisation is the origin of coronal mass ejections (CMEs). Once initiated, CMEs can dominate the inner heliosphere, and stellar CMEs are likely to be a major influence on exoplanetary atmospheres.

\begin{figure}[t]
{\includegraphics[width=\textwidth]{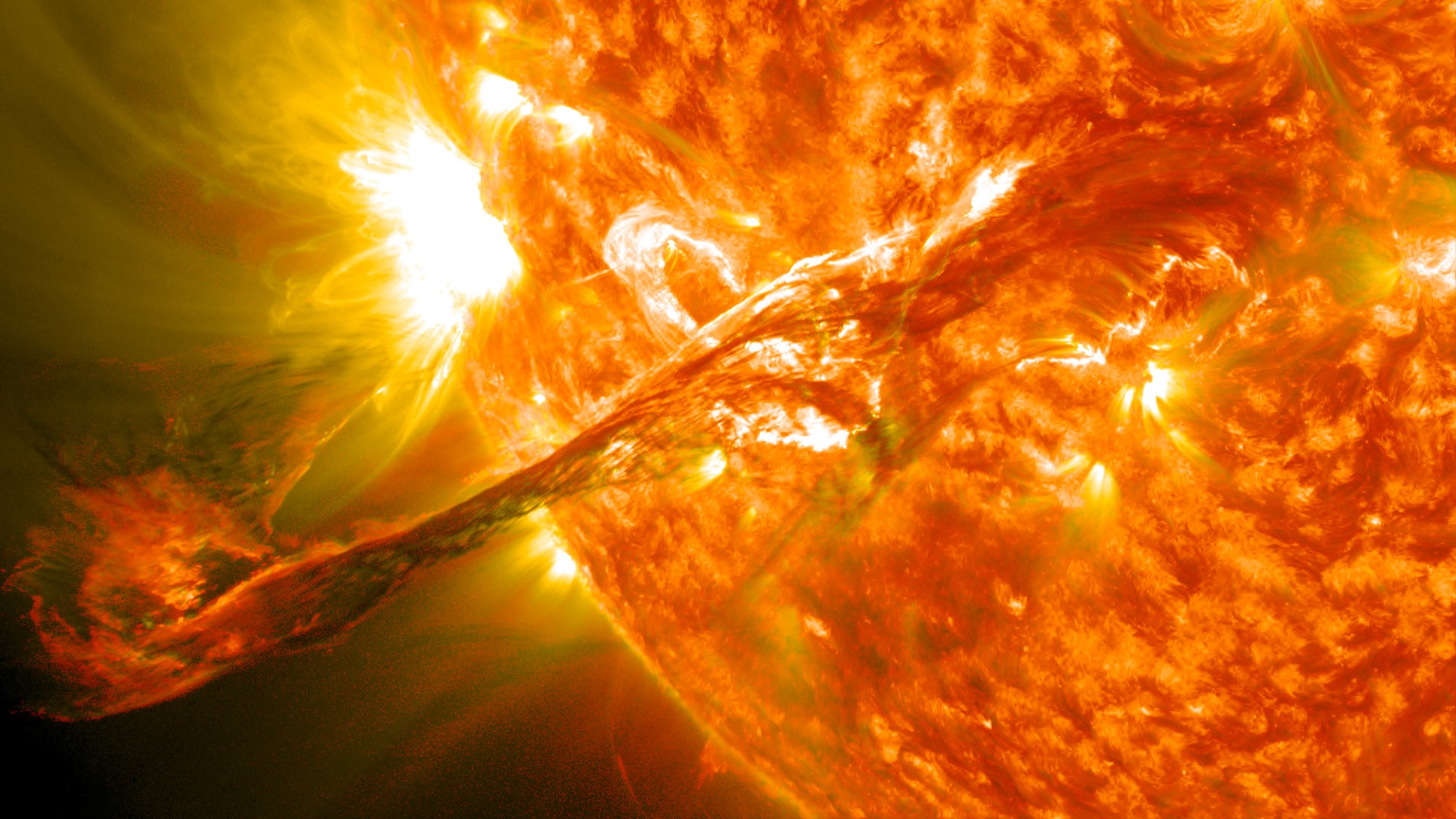}}
\caption{%
Erupting prominence on the Sun on 31.\,8.\,2012 at 19:49\,UT. This shows a composite of plasma at 10$^{\,5}$\,K (red) and 10$^{\,6}$\,K (yellow) erupting into interplanetary space. The complex internal structure is either a remnant of the precursor structure in the low corona or forms during the eruption. Direct measurements of the magnetic field are essential to distinguish between these and other scenarios. 
Image credit: NASA's GSFC and SDO/AIA Team (https://apod.nasa.gov/apod/ap180916.html)
\label{F:eruption}
}
\end{figure}

A CME expands when it lifts off. In this process, some CMEs reveal their complex internal structure (e.g.\ \fig{F:eruption}). Our current information is based on intensity images, therefore no direct information on the magnetic field is available.
Only direct measurements of the magnetic field will allow us to distinguish between the very different physical mechanisms that have been proposed for the initiation of CMEs \cite[][]{2011LRSP....8....1C}. This knowledge is needed not only to strongly improve the predictability of CMEs and their space weather effects, but also to predict what the CMEs on other stars could look like \cite[e.g.][]{2018ApJ...862...93A,2019NatAs.tmp..328A}.
Coronagraphic direct measurements of the magnetic field in the corona will close this gap in our knowledge.

Many  erupting features show a high degree of magnetic twist and writhe. Magnetic helicity quantifies how elementary magnetic flux tubes are wound around each other in a defined volume. In a closed system, magnetic helicity is almost conserved in resistive magnetohydrodynamics (MHD) on~timescales shorter than the global diffusion timescale \cite[][]{Matthaeus1982, Berger1984}.
Magnetic helicity can be used to quantitatively link flux ropes that are observed in situ in the solar wind to their sources in the solar atmosphere. Magnetic helicity also constrains the dynamical evolution of the ejected magnetic field in the solar wind \cite[e.g.][]{2009JGRA..114.2109D}.
Determining the magnetic helicity carried by a CME requires measuring the magnetic field and is therefore fundamental to understanding the evolution of CMEs.

The magnetic field is reconfigured on large scales in the course of large-scale disruptions.
This magnetic reconfiguration requires that the connection to the small-scale magnetic features in the lower atmosphere changes as well. This is illustrated by the emergence and submergence of magnetic flux on the solar surface and the tether-cutting mechanism for CME eruption. Information on the magnetic field beyond what extrapolations can provide is required to understand the physics of eruptions. Extrapolations of the magnetic field from the surface level are currently used to investigate the change in free energy of the magnetic field. However, as mentioned above, before and during dynamic events the assumptions of these extrapolations break down, . and it is mandatory to directly measure the coronal magnetic field. This underlines the need for direct coronagraphic measurements of the magnetic field during eruptions.

The most spectacular events on the Sun are flares and CMEs. They govern the space weather. During such events, particles are accelerated to high energies and can leave the Sun. If they hit Earth, they cause a wide range of effects, including damage to satellites and harm to astronauts. With an understanding of the magnetic processes before and during the initiation of flares and CMEs, we will finally be able to understand the roots of space weather.

\subsection{High-energy processes in the atmosphere}

Particles are accelerated to high energies during flares and CMEs. Energetic processes are not limited to large scales, however. They essentially occur on all scales, including those we cannot currently resolve. For example, it has been proposed that particles are also accelerated during unresolved energy release events (nanoflares), but we have only indirect evidence for this \cite[][]{2014Sci...346B.315T}. These nanoflares are associated with magnetic reconnection. Measuring the magnetic configuration and its change during such events will greatly advance our understanding of the basic plasma physics process of reconnection and particle acceleration.

Our diagnostics for reconnection in the upper solar atmosphere are currently restricted to imaging the thermal response of the plasma that is associated with the heating during such events, in particular through extreme-UV imaging \cite[e.g.][]{2014ApJ...787L..27S,2018ApJ...868L..33L}. The presence of current sheets is often indirectly derived from thin threads that are  visible in extreme-UV images. For a few cases, however, observations have revealed current sheets through direct measurements of the magnetic field. They were all very low in the atmosphere \cite[][]{2003Natur.425..692S,2004A&A...414.1109L}.

Measuring the change in the magnetic field in space and time around the reconnection site is very important to understand the heating and particle acceleration at a reconnection site. This alone gives access to the structure of the current sheets that form during the event and to the location and manner in which particles are accelerated. Imaging observations in the extreme UV and X-rays can show the changing structure of the magnetic field (assuming that the emission indeed outlines the magnetic field). We can quantify how much magnetic energy is converted into other forms of energy only through directly measuring the magnetic field because some of that energy is in forms that are hard to detect (e.g. in the form of energetic particles, or in small brightenings in the visible and IR wavelength ranges).
Measuring the magnetic field will therefore be essential to derive the energy that is available for particle acceleration and for the generation of (hard) X-ray emission during the flare events.

\subsection{Relation to other science fields}

The upper solar atmosphere is the source region of the radiation in the UV. This radiation is of paramount importance for Earth's atmosphere. It controls the chemistry and heating in the stratosphere and in higher layers of the atmosphere of our home planet, and thereby may significantly affect the long-term climate evolution. The extreme-UV radiation from the corona is important for the upper atmosphere of Earth, and the near-UV radiation affects the stratosphere. The near-UV originates from the solar photosphere and chromosphere. To date, the relative variability in the UV radiation and its contribution to solar variability is not fully understood, although it is clear that it is driven by magnetic field variations.  When we understand why the outer atmosphere becomes unstable, we can scientifically define the ingredients of space weather, which through energetic particles and CMEs impacts the terrestrial environment.

The science proposed here is also relevant for host stars of other planets. 
Far more violent disruptions than on the Sun have been observed in the outer atmospheres of other stars. For example, super-flares have been found on sun-like stars \cite[e.g.][]{2000ApJ...529.1026S,2012Natur.485..478M}. The energy released by these flares exceeds the energy released by the largest flares on our Sun by 1000 times. Stars that are more active than the Sun can produce CMEs that then propagate into the astrosphere of the star \cite[e.g.][]{2019NatAs.tmp..328A}.
It is clear that this will affect the exoplanets of these stars in terms of their (magnetic) environment and possible habitability.

Understanding the magnetic processes in the outer solar atmosphere will also greatly enhance our understanding of accretion disc coronae and winds. We can expect the same processes to operate in these magnetically dominated environments as in the solar corona. As in many other cases, the Sun can serve here as a template and laboratory for other astrophysical objects.

The science we propose will also contribute to basic space plasma physics (e.g. current sheets, reconnection, and instabilities). By imaging the magnetic field in the outer solar atmosphere, we can characterise the large- and small-scale structure. This is otherwise only possible through in situ measurements in planetary magnetospheres (and possibly the heliosphere) with large swarms of probes. We can study the cross-scale coupling through one single solar observatory as proposed here.

\section{Measuring magnetic fields by remote-sensing methods\label{S:method}}

Almost all the information on the magnetic field is contained in the polarisation profiles 
of solar spectral lines. In other words, it is contained in the wavelength variation of the 
Stokes parameters $Q$ and $U$ (which quantify the linear polarisation) and of
the Stokes parameter $V$ (which quantifies the circular polarisation). 
Together with the Stokes parameter $I$ (i.e. the intensity), these Stokes 
parameters fully characterise any beam of electromagnetic radiation. 
In order to carry out quantitative investigations of 
the magnetic fields in the solar atmosphere, we 
therefore require telescopes that are equipped with instruments capable of measuring 
the four Stokes parameters in suitable lines of the solar electromagnetic 
spectrum. The relevant physical mechanisms that produce polarisation in solar spectral 
lines are described next.

\subsection{The Zeeman effect}

In the presence of a magnetic field, the spectral lines split into 
various components with specific polarisation properties that depend on the 
angle between the line of sight and the magnetic field vector. 
While the circular polarisation signal (caused by the longitudinal component of 
the magnetic field) scales with the ratio ${\cal R}$ between the Zeeman 
splitting and the Doppler width, the linear polarisation signal (caused by
the transverse component of the magnetic field) scales as ${\cal R}^2$. 
The Zeeman 
effect is therefore more prominent in circular polarisation than in linear 
polarisation.  
Moreover, because $\mathcal{R}$ increases with the spectral line 
wavelength, the Zeeman effect is much more effective in the IR than in the UV.
This makes it a valuable tool for coronagraphic observations above the limb in 
IR lines of highly ionised species, such as 
Fe~{\sc xiii} at 1074~nm \cite[][]{2000ApJ...541L..83L}.
The circular polarisation  can also be measured in some UV lines from the upper solar 
chromosphere. 
This has been demonstrated by the 
CLASP-2 suborbital rocket experiment. With a telescope with an aperture of only 27~cm, it could measure the 
circular polarisation produced by the Zeeman effect in the Mg~{\sc ii} h \& k 
resonance lines in a plage region.

\subsection{Linear polarisation by scattering processes}

The scattering of anisotropic radiation produces linear polarisation in a 
multitude of solar spectral lines, without the need of any magnetic field. 
This especially occurs in many of the spectral lines that originate in the 
outer solar atmosphere, where the anisotropy of the 
radiation field coming from the underlying solar disc is higher than in the
photosphere.
The physical origin of this 
resonance-line polarisation lies in the fact that the radiative transitions 
induced by the incident anisotropic radiation produces atomic level 
polarisation, that is, population imbalances and quantum coherence among the 
magnetic sublevels of degenerate atomic energy levels. This in turn 
produces polarisation in the radiation that is re-emitted  
in the scattering process.

\subsection{The Hanle effect}

The atomic level polarisation
is induced by radiation. This polarisation is modified through the magnetic
field; this is called the Hanle effect. Therefore, the Hanle effect also modifies the
linear polarisation that is produced by the scattering of anisotropic radiation. For the Hanle effect, the degeneracy of the levels is lifted in the presence of a magnetic field, and the radiation-induced atomic polarisation is modified. 
The characteristic magnetic field dependence of the linear polarisation of the scattered spectral line radiation that is caused thereby is suitable for exploring the outer solar atmosphere.
Importantly, the Hanle effect is sensitive even for magnetic fields that are tangled at sub-resolution scales \citep[e.g.][]{2004Natur.430..326T}.

There is a critical magnetic field strength ($B_H$) for the onset of the Hanle effect. This is independent of the plasma temperature and of the wavelength of the considered spectral line.
Typically, measurements through the Hanle effect are sensitive in a range of $0.1\,B_H$ to $10\,B_H$ of the critical value.
For magnetic fields that are significantly above the critical value $B\,{>}10\,B_H$, the line is saturated. Then the linear polarisation of the emerging spectral line radiation is sensitive only to the orientation of the magnetic field.
The expected magnetic field in the source region of the observation thus determines the choice of lines with different critical strengths $B_H$.

\begin{figure}[t]
\centerline{\includegraphics[width=0.7\textwidth]{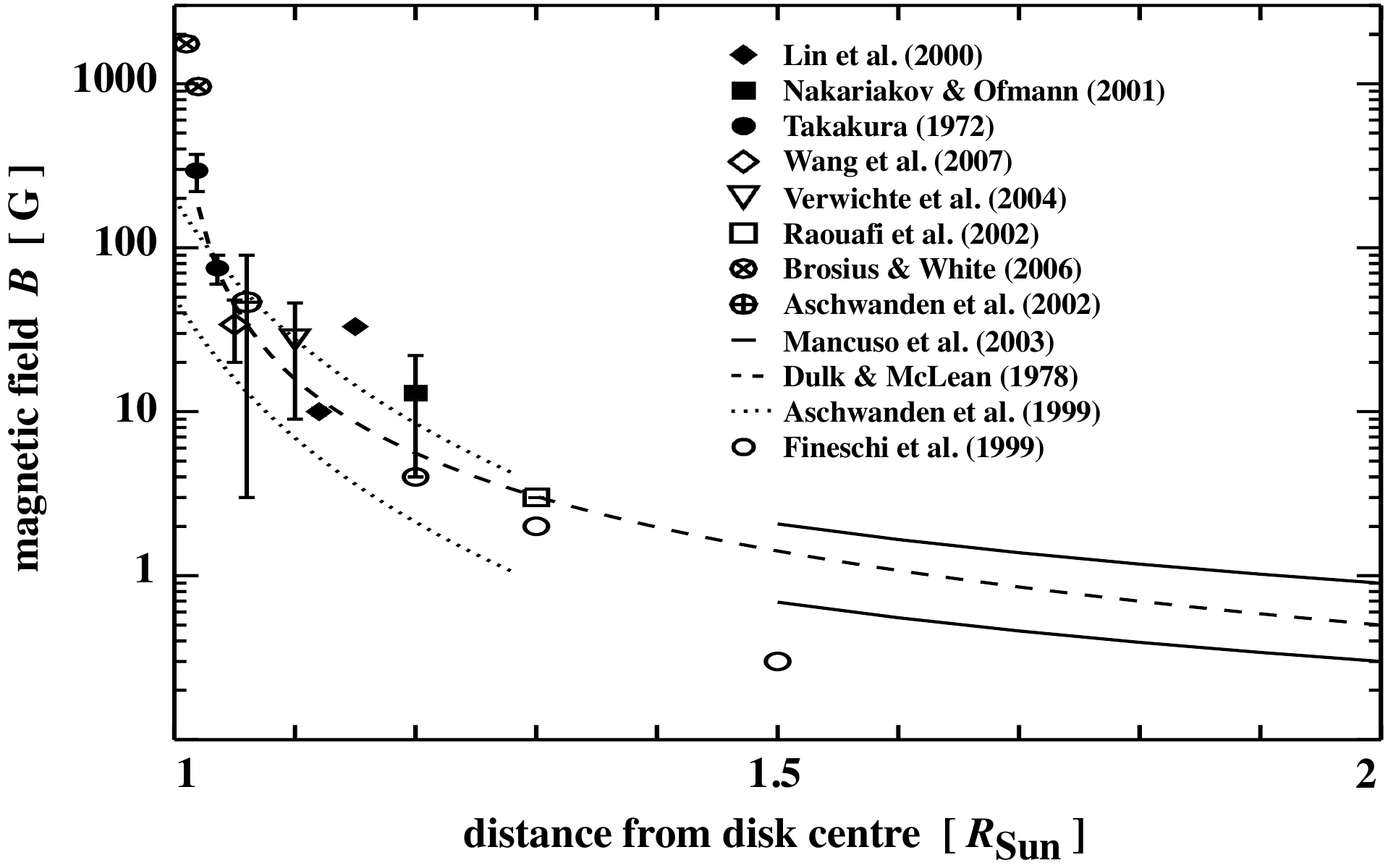}}
\caption{%
Expected magnetic field strength $B$ in the outer solar atmosphere.
This represents the average drop of the magnetic field from the surface (at one solar radius) out into the corona.
Large spatial variations are expected.
Compiled from different sources
\cite[based on][]{2012ExA....33..271P}.
\label{F:field}
}
\end{figure}

The magnetic field on the Sun drops dramatically from the surface to the outer atmosphere. The field at the surface reaches several 1000\,G, in the chromosphere we expect a few 100\,G, and the field in the corona drops to 1\,G and below (see \fig{F:field}).
The Ly-$\alpha$ line of \ion{H}{1} at 121\,nm is ideally suited for diagnostics of the chromosphere on the disc and for coronagraphic observations
of the corona above the limb. Its critical Hanle value is $B_H = 53$\,G.
\ion{The Mg}{2} line at 280\,nm ($B_H = 22$\,G) is also well suited, but it  is only visible on the disc.
The critical values of IR lines for coronagraphic observations  (e.g.\ \ion{Fe}{13} at 1075\,nm) are well below 1\,G. This means that they will be in saturation and can only provide the field orientation.
Extreme UV permitted lines for off-limb and on-disk observations (e.g.\ \ion{Fe}{10} at 17.4\,nm) will only provide information on the field orientation (apart from the electron density) because their critical magnetic field for the onset of the Hanle effect is much larger than the magnetic strength expected for the solar coronal plasma \cite[][]{2009ASPC..405..423M}.

\subsection{Magneto-optical effects in strong resonance lines}

In strong resonance lines such as \ion{H}{1} Ly-$\alpha$ and Mg~{\sc ii} h \& k, 
the joint action of quantum interference between the sub-levels pertaining to
the upper $J$ levels of these lines and the correlation effects between the 
incoming and outgoing photons in the scattering events (partial frequency 
redistribution) produce very strong $Q/I$ signals in the wings of these lines
\citep[see][]{2012ApJ...750L..11B,2012ApJ...755L...2B}.

Without the need of any magnetic field the value of $Q/I$ can be very large in the wings of strong resonance lines.
The magneto-optical terms of the Stokes-vector transfer equation then introduce magnetic sensitivity in the wings of $Q/I$ and $U/I$.
This happens already for magnetic fields with strengths similar to those needed for the onset of the Hanle effect. 
Such magneto-optical effects are expected to produce 
measurable magnetic sensitivity in the wings of the Mg~{\sc ii} h \& k lines 
\citep{2016ApJ...831L..15A,2016ApJ...830L..24D} and in the wings
of the \ion{H}{1} Ly-$\alpha$ line 
\citep{2019ApJ...880...85A}.

To summarise, the polarisation signals in UV and IR spectral lines are sensitive to the presence of magnetic fields throughout the whole outer atmosphere.
The Hanle effect is sensitive to the (weak) magnetic fields expected for the upper chromosphere and transition region because it operates at the line centre where the opacity is high. The magneto-optical effects that operate in the wings of strong resonance lines are sensitive to the presence of similarly weak magnetic fields in the underlying chromospheric and photospheric layers because  the opacity in the line wings is lower. 
The circular polarisation produced by the Zeeman effect in some UV lines (e.g.\ Mg {\sc ii} at 280\,nm) is sensitive to the longitudinal component of the magnetic field in the solar chromosphere.
The linear polarisation produced by scattering in UV lines has been successfully exploited to investigate the solar coronal magnetic field \citep[e.g.][]{2002A&A...390..691R,2002A&A...396.1019R}.
From the upper chromosphere and transition region into the corona, the two sub-orbital rocket flights of CLASP obtained and exploited spectropolarimetric data in the Ly-$\alpha$ line at 121\,nm \cite[][]{2017ApJ...839L..10K,2018ApJ...866L..15T} and \ion{Mg}{2} near 280\,nm.
In permitted (electric dipole) UV lines and in forbidden (magnetic dipole) IR lines, measurements of the polarisation produced by atoms in the solar coronal plasma can probe the magnetism of solar coronal structures, and the solar wind velocity \citep[e.g.][]{2017SSRv..210..145C,2017SSRv..210..183T}.

When spectropolarimetric measurements in the UV are complemented by IR observations, new discoveries in solar and stellar physics are facilitated: through the Hanle and Zeeman effects, the \ion{He}{1} triplet at 1083\,nm is sensitive to the magnetic field of plasma structures in the chromosphere and corona. 
UV lines such as  Ly-$\alpha$ at 121\,nm and the resonance and subordinate lines of Mg {\sc ii} (all around 280\,nm) are suitable for probing the upper chromosphere.

A space telescope that is equipped with instruments that are designed to measure the four Stokes parameters in suitably chosen UV and IR spectral lines will be able to achieve a true revolution in our empirical understanding of solar magnetism.

\section{Integral role of realistic atmospheric models\label{S:models}}

The polarisation measurements
need to be inverted to derive the magnetic field from observations.
More than a century ago, magnetic fields on the Sun were discovered through the Zeeman splitting of spectral lines in the strong magnetic field of a sunspot \cite[][]{1908ApJ....28..315H}. 
This was possible even without analysing the polarisation state of the light. When the geometry of the magnetic field or the thermal structure of the atmosphere is more complex, the full information of the polarisation has to be taken into account to reconstruct the full 3D vector of the magnetic field, i.e.\ its strength and direction.
Not only the Zeeman effect, but also the Hanle effect has to be considered for the weaker magnetic fields in the outer atmosphere. 
This requires an inversion of the observed polarisation data.

Several inversion codes are available for the chromosphere that include the Zeeman effect \cite[][]{2000ApJ...530..977S,2019A&A...623A..74D}. In the next few years, they will be extended to also cover the Hanle effect, so that all available types of data can then in principle be invertible. Initially, these inversions will  be made in 1D, but will include non-LTE effects (i.e.\ departures from local thermodynamic equilibrium). Ultimately, 3D versions will be required, and first steps in this direction have been taken.~Inversion codes have  been developed and implemented that  take into account the coupling of neighbouring pixels in the observation \cite[][]{2012A&A...548A...5V,2013A&A...557A..24V}. This coupling is introduced by the point spread function of the instrument.

\begin{figure}
\centerline{\includegraphics[width=\textwidth]{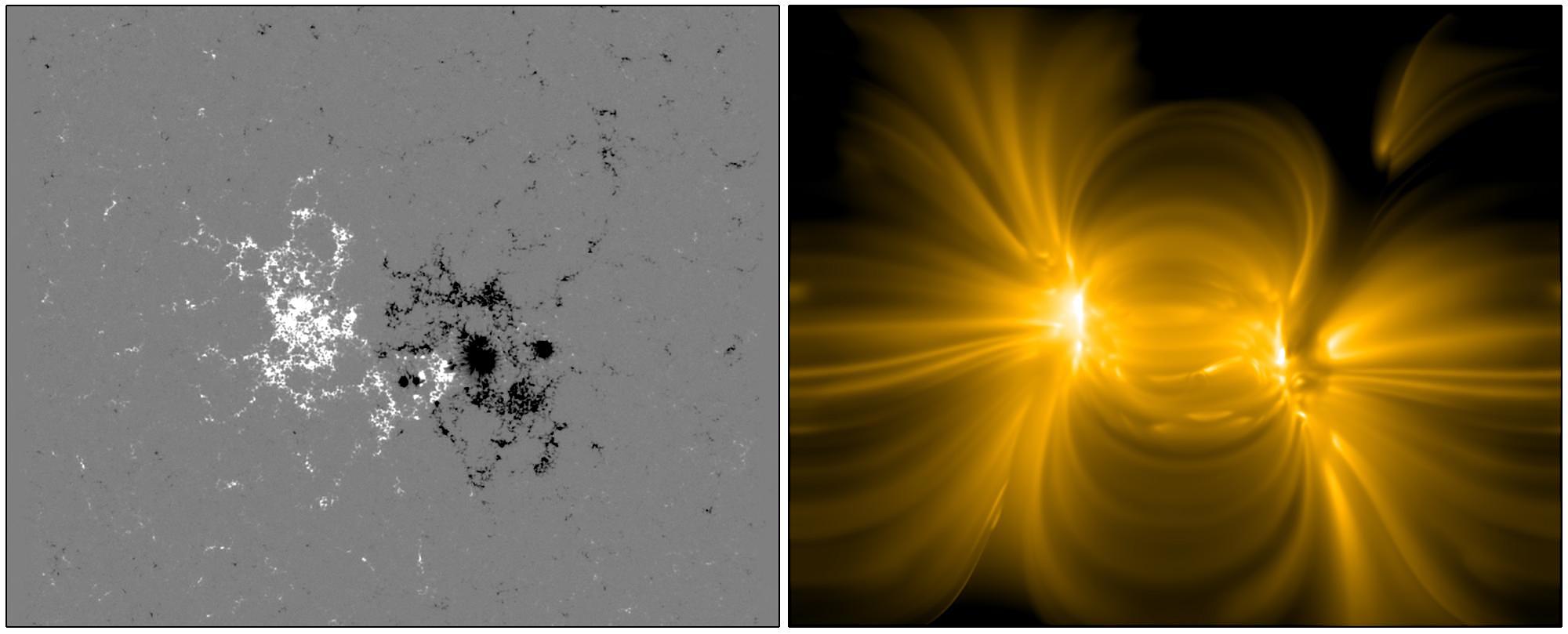}}
\caption{Snapshot from a data-driven 3D MHD model of an active region. 
The model uses the evolving observed photospheric magnetic field to drive the evolution of the solar atmosphere from the photosphere all the way into the corona. 
The left panel shows the vertical magnetic field in the photosphere (in the range of $\pm$500\,G) and the right panel the coronal emission as it would be observed by the 171\,\AA\ channel of AIA showing ca.\ 10$^6$\,K hot plasma when looking from straight above, i.e.\ close to disk center.
The panels show a 375$\times$300\,Mm$^2$ part of the computational domain.
The model captures the large-scale structure of the corona above the modelled active region (AR 12139 from 16. Aug. 2014) quite well.
Based on \cite{2019A&A...624L..12W}.
\label{F:activeregion}}
\end{figure}

A very important role will be played by realistic models of the outer solar
atmosphere. Here a numerical simulation that includes all relevant physical processes
provides the evolution of the plasma and the magnetic field in a 3D volume.
A forward-modelling code is then used to calculate the emission, including
the state of polarisation, from the simulation results (see \fig{F:activeregion} for an example). The numerical simulations
are currently mostly run by solving the 3D MHD\ equations (with extensions).
Examples are MURaM \cite[][]{2017ApJ...834...10R} and Bifrost \cite[][]{2016A&A...585A...4C}, both including the radiative transfer problem, or Pencil that allows to model a data-driven corona \cite[][]{2019A&A...624L..12W}.
Several 3D non-LTE codes also exist for the forward modelling of the radiation in spectral lines, e.g.\ Multi-3D
\cite[][]{2009ASPC..415...87L} for the intensity and PORTA \cite[][]{2013A&A...557A.143S} for all Stokes parameters. 
This shows that the basic tools are available even now so that numerical models can be used to
evaluate and improve inversion techniques with which the magnetic field can be derived from
observations.

Numerical simulations of the outer solar atmosphere are computationally demanding.
They can be included in inversions of spectropolarimetric observations within a few years, however, and first steps have already been taken \cite[][]{2017ApJS..229...16R}.
For example, forward-modelling calculations today
provide an essential reference to identify and distinguish the impact of various physical processes that affect scattering polarisation.
This approach is necessarily model dependent, but the availability of increasingly realistic MHD simulations  make it the most reliable approach to infer information about the magnetism of the outer solar atmosphere.

Realistic atmospheric models allow us to investigate the parameter space of
possible polarisation signals and thus can show new paths for interpreting the
data. In the recent IRIS mission \cite[][]{2014SoPh..289.2733D}, for instance,
numerical models have been an integral part of the data analysis. By exploring
the spectral line profiles that were found in the forward models,
the profiles could be classified and were used to derive physical quantities from features
in the \ion{Mg}{2} h and k line profiles \cite[e.g.][]{2013ApJ...778..143P}.
Future realistic 3D models will play the same role for interpreting polarisation data from the outer solar atmosphere.
Models like this will in particular be essential for understanding how the different
spectral lines form in different regions (heights) of the outer atmosphere
and how the different lines can be best combined for a reliable inversion of the
magnetic field.

We proceed with a brief overview of models for the chromosphere
and (large-scale) corona. We also consider future developments.

\subsection{Chromospheric models}

Simulations of the solar chromosphere require challenging theory: the
energy densities of magnetic field, plasma, and waves are all of a similar
magnitude, which
provides a wealth of dynamics. Radiation cannot be treated as
optically thin as in the case of the corona, nor can the atoms be assumed to be in local thermodynamic equilibrium (LTE)\ as in the photosphere. The gas in the chromosphere is dynamic and only partially ionised, so that additional effects have to be taken into account, such as ambipolar diffusion and time-dependent partial ionisation. 

The past decade saw a very fast evolution of chromosphere modelling, and
different models can now account for a range of the required physics,  including
non-equilibrium populations \citep{2007A&A...473..625L,2016ApJ...817..125G},
non-LTE radiation transfer \citep{2012A&A...539A..39C},
two-fluid MHD \citep{2019A&A...627A..25P}, and non-ideal
MHD \citep{2012ApJ...753..161M,2018A&A...618A..87K}.
With further progress at this pace, we can expect to have reasonably realistic 3D numerical
experiments of the solar chromosphere in another decade. In other words, the required models
will be available when they are needed to develop and evaluate improved inversions
to retrieve the chromospheric magnetic field.

\subsection{Coronal models}

Numerical models for the large-scale corona potentially involve fewer physical
processes than are required in the chromosphere. The coupling across
scales and the loss of collisional coupling between particles adds to the
problem, however. The former requires high spatial resolution or adaptive grids, the
latter means that kinetic models or particle-in-cell simulations are necessary.
These effects have been considered, but are still far from being realistic in
the sense of reproducing actual coronal structures. Still, significant progress
has been achieved, e.g.\ in modelling the dynamics of flares \cite[][]{2019NatAs...3..160C}.

Another complication in the large-scale corona is the integration along the
line of sight of the optically thin emission that prevents simple direct
inversions of the magnetic field \citep[][]{Kramar2006,Kramar2013}. Recent
forward models show paths to interpreting coronagraphic measurements by minimising
the difference between synthesised coronal 
observables and real coronal data \citep[e.g.][]{Dalmasse2019}.
In the same way as for the chromospheric models, we can expect (global) coronal models to be ready within a
decade to assist in the inversion of the polarimetric measurements of the
global corona.

\subsection{Outlook}

When realistic 3D models of the chromosphere and the corona become available,
the assimilation of data into the models will be a powerful tool to better
understand the structure of the outer solar atmosphere and its evolution.
The experience of data assimilation in weather forecasting shows the enormous
potential of this technique. Continuously assimilating the acquired data
into the model provides a way to 
iterate from the measured polarisation maps towards a correct solution of
the magnetic field structure and its evolution.
These are ensemble models, therefore the computations require  two orders of magnitude
more computational power than current models. With more efficient codes
and the expected progress of computational resources in the next decade, however, this will become feasible. 
This would then provide an advanced tool for analysing the polarimetric measurements
of the outer solar atmosphere.

\section{Scientific requirements for measuring the magnetic field in the
upper solar atmosphere\label{S:requirements}}

\subsection{Limitations in current approaches\label{S:limits}}

Measurements of the magnetic field at the surface of the Sun (the photosphere) are currently routinely performed at high spatial resolution in smaller regions, and with moderate resolution when they cover the whole Sun. This is achieved by remote-sensing observations, mainly through the Zeeman effect (see Sect.\,\ref{S:method}). However, the magnetic field at only about 1000\,km above the surface dominates the gas in terms of energy, i.e. plasma-$\beta$ becomes smaller than unity. To understand how the magnetic field governs solar activity above the surface (in the chromosphere and corona), we require reliable and routine measurements there as well.

To estimate the magnetic field in the outer solar atmosphere, we currently
mostly have to take recourse to extrapolations of the magnetic field based on observations
of the solar surface. While this has reached a considerable degree of sophistication,
these methods are not satisfactory because there are limitations in principle \cite[e.g.][]{2009ApJ...696.1780D}. Also, it is not possible to capture the temporal
evolution of dynamic phenomena from the chromosphere to the atmosphere at
large with extrapolation methods that assume a static or
stationary situation. This prevents further progress in solar atmospheric
research because the magnetic field structures and drives the
dynamics in the solar atmosphere, from the chromosphere to the outer corona.
Therefore we need direct observations of the magnetic field not only in the photosphere, but throughout the whole outer solar atmosphere.

With the instrumentation we have so far, attempts to measure the magnetic
field in the corona have been more a proof of concept than regular observations.
The first of such observations above the limb using a coronagraphic technique
in the IR had very low spatial resolution and required very long exposure
times \cite[][]{2000ApJ...541L..83L,2004ApJ...613L.177L}. While the field
progressed, observations are still sparse and limited to a few sample structures
on the Sun. For example, a post-flare-loop system above the limb
has recently been observed in some detail in the IR 
\cite[][]{2019ApJ...874..126K} and in radio wavelengths \cite[][]{2018ApJ...863...83G}. This
gave consistent results.

Observations of the magnetic field in the chromosphere have been obtained
in a more regular fashion, but are mostly limited to the diagnostics in the
lines of \ion{He}{1} at 1083\,nm and the triplet of \ion{Ca}{2} around 854\,nm.
Observations in these lines are limited in that they do not provide information on the upper chromosphere, where the magnetic field energetically dominates the plasma. Observations in the UV part of the spectrum are required to reach the upper chromosphere and even higher layers.

Previous observations on sub-orbital rockets showed the potential of the polarisation that is measured in the UV in Ly-$\alpha$ and Mg\,II for the magnetic field measurements in the chromosphere (see Sect.\,\ref{S:method}). Polarisation measurements have also been performed for C\,IV by UVSP on SMM from space \cite[][]{1982SoPh...81..231H,1983SoPh...84...13H} for the hotter layer of the transition region to the corona. The short rocket observation in Ly-$\alpha$ and Mg\,II only provided a glimpse of polarisation measurements in the UV, however, and the existing measurements in C\,IV are not reliable \cite[][]{2001ASPC..248..553L}.

Upcoming and planned space-based solar observatories will not provide polarimetric information in the UV. They will therefore not provide measurements of the magnetic field in the uppermost layers of the chromosphere. We propose to close this observational gap in this paper.

\subsection{Scientific requirements\label{S:sci.requirements}}

To measure the magnetic field in the upper atmosphere, we distinguish (a) measurements in front of the solar disc and (b) measurements above the limb. Observations on the disc will give access to the small-scale structures in the chromosphere that form the base of the outer atmosphere and thereby shape the magnetic connection throughout the whole atmosphere (see Sect.\,\ref{S:req:chromo}). The off-disc measurements will provide information on the large-scale structure of the magnetic field, giving a view of the corona similar to what is seen during a solar eclipse (see Sect.\,\ref{S:req:corona}). Finally, the small-scale chromospheric features have to be connected to the global corona, for which observations in the extreme UV are crucial because they can also give access to coronal structures when observations are made on the solar disc (see Sect.\,\ref{S:req:connect}). To describe the respective requirements, we restrict the discussion to one example for each case. More than one process that constrains each requirement in most cases. The scientific requirements for the respective observations are summarised in Table\,\ref{tab:sci.req}.

\begin{table*}
\begin{center}
    \caption{%
      Overview of scientific requirements.%
      \label{tab:sci.req}}
    \begin{tabular}{@{}l@{~}ccc@{}}
    \hline\hline
    & chromospheric         &                   & extreme UV \\
    & telescope             & coronagraph       & coronal imager \\
                                & \small (Sect.\,\ref{S:req:chromo}) 
                                & \small (Sect.\,\ref{S:req:corona}) 
                                & \small (Sect.\,\ref{S:req:connect}) \\
    \hline
    Spatial resolution          & 0.1\arcsec    
                                & 2 -- 10\arcsec~$^{(b)}$ & 1.2\arcsec
    \\
    Temporal resolution         & 1\,s         
                                & 5 -- 100\,s~$^{(b)}$    
                                & 3 -- 60\,s~$^{(e)}$
    \\
    Field of view               & 300\arcsec$\times$300\arcsec 
                                & 1.1 -- 3.0\,$R_\odot$~$^{(c)}$
                                & 0.8$^\circ${$\times$}0.8$^\circ$
    \\
    Spectral resolution or band width  & $\lambda/\Delta\lambda$: 30\,000 
                                & $\Delta\lambda$: 5\,nm (UV); ~ 0.1\,nm (IR)
                                & $\Delta\lambda$: 0.35\,nm
    \\
    Signal-to-noise ratio~$^{(a)}$
                                & $3\times10^3$
                                & $10^2$
                                & $10^3$
    \\
    \hline
    Key spectral lines/ranges   & \ion{Mg}{2} (280\,nm)     & VL (pB)~$^{(d)}$ & \ion{Fe}{10} (17.4\,nm)\\
    \small (small selection only) & \ion{C}{4} (155\,nm)      & Ly-$\alpha$ (121\,nm)\\
                                & \ion{H}{1} Ly-$\alpha$ (121\,nm) & \ion{Fe}{13} (1075\,nm) \\
                                & \ion{Ca}{2} IR (854\,nm)  & \ion{He}{1} (1083\,nm) \\
                                & \ion{He}{1} (1083\,nm)    &  \\
    \hline
    \end{tabular}
    \\[-1.5ex]
    {\small
    \begin{tabbing}
    {\bfseries{Notes:~~}}
    \= $^{(a)}$~Ratio of the signal in the line compared to the noise.
    \\
    \> $^{(b)}$~Depending on observation mode: polarised brightness, intensity only, or polarisation.
    \\
    \> $^{(c)}$~Radial distance from disc centre, i.e. from 0.1 -- 2.0\,$R_\odot$ above the limb.
    \\
    \> $^{(d)}$~Visible light (polarised brightness), broad-band observation in the visible.
    \\
    \> $^{(e)}$~Faster cadence for imaging-only mode, slower cadence for polarimetric observations.
    \end{tabbing}
    }
\end{center}
\end{table*}

\subsubsection{Magnetic field in the chromosphere\label{S:req:chromo}}

\begin{figure*}[t]
{\includegraphics[width=\textwidth]{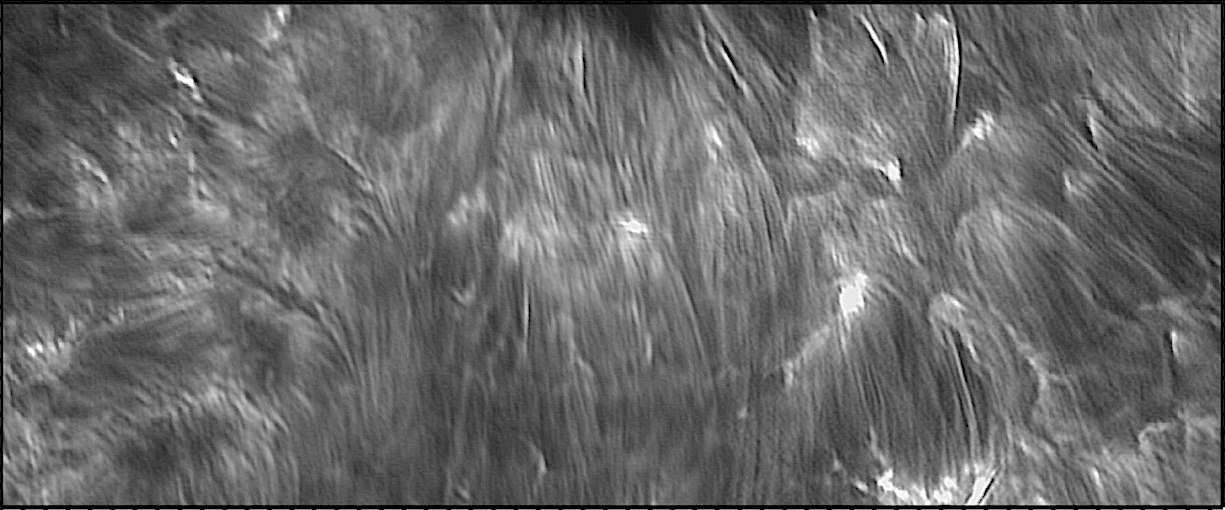}}
\caption{%
Fibrils seen in the solar chromosphere in \ion{Ca}{2} K.
The field of view is only 34\arcsec$\times$14\arcsec\ and thus tiny compared to the solar diameter of about 0.5$^{\,\circ}$ \cite[based on][]{2017ApJS..229...11J}.
\vspace{-2.5ex}
\label{F:fibrils}
}
\end{figure*}

The chromosphere is a highly structured and very dynamic part of the solar atmosphere. Current 1m class ground-based and balloon-borne solar telescopes show features down to scales smaller than 100\,km. The Sunrise telescope, for example, revealed fibrils (\fig{F:fibrils}) with widths ranging from 100 to 270\,km \cite[][]{2017ApJS..229....6G}. These fibrils are generally thought to map the magnetic field \cite[][]{2017ApJS..229...11J} and are heated when magnetic flux emerges \cite[][]{2018A&A...612A..28L}.
Current instruments do not allow us to directly measure the magnetic field in these fibrils, and similar conclusions apply to other key chromospheric features.
This leads to the requirement for an {{angular resolution of 0.1\arcsec}}\ for chromospheric magnetic measurements, which corresponds to about 70 km. Only by achieving this resolution will we be able to reveal the formation and evolution physics of these chromospheric structures.

The required temporal resolution  is closely linked to the spatial resolution. Essentially, the time resolution should be equivalent to the time it takes for a perturbation to cross the resolution element. The relevant speed for magnetic changes is the Alfv\'en velocity, which is of the order of 10 to 100 km/s in the chromosphere. Consequently, the observations have to allow for a {{temporal resolution of about 1\,s}}.

Because of the cross-scale coupling, the field of view of the chromospheric observations has to encompass a whole active region on the Sun. This applies to the observations of the chromosphere for its own sake and for the magnetic connection through the atmospheric layers. For example, magnetic flux often emerges and cancels on small scales \cite[][]{2017ApJS..229....4C}, but it might still govern the formation and evolution of hot structures in the core of an active region \cite[][]{2018A&A...615L...9C}. The required {{field of view is therefore 300\arcsec$\times$300\arcsec}} to fully cover an active region. This corresponds to 3\% of the visible solar disc.

The spectral resolution for spectroscopic and spectropolarimetric observations is set by the expected velocity (dispersion) seen in the spectral lines and by the need to resolve the polarisation signals, which are often narrower than the intensity profiles of the spectral lines. The expected velocities in the chromosphere are of the order of 1\,km/s and higher, as we know from existing spectroscopic observations. A spectral resolution corresponding to 10\,km/s is sufficient to determine the shift of an optically
thin line (through centroiding). This corresponds to a {{spectral resolving power $\lambda/\Delta\lambda$ of 30\,000}}.
To resolve features in optically thick chromospheric lines, such as self-absorption in \ion{Mg}{2}, an equivalent spectral resolution is required, as we know\ from observations with IRIS \cite[][]{2014SoPh..289.2733D}.

Line selection is a fundamental question for a spectropolarimetric instrument. Chromospheric lines  investigated from the ground for magnetic measurements are mostly from \ion{Ca}{2} and \ion{He}{1}. The IR triplet of \ion{Ca}{2} 854\,nm has been used for quiet Sun and active regions \cite[e.g.][]{2019A&A...621A...1R}, while no robust routine observations have been possible in the H and K lines of \ion{Ca}{2} so far. The \ion{He}{1} line at 1083\,nm has been widely used for magnetic diagnostics \cite[e.g.][]{2003Natur.425..692S}, even though it is mostly limited to active regions. The \ion{He}{1} D$_3$ line at 588\,nm has also recently been evaluated during a flare \cite[][]{2019A&A...621A..35L}.
To interpret the observations, it is important to perform multi-line inversions. This has been amply demonstrated for photospheric observations \cite[][]{2019A&A...622A..36R}. To apply such techniques to the chromosphere, two of the most prominent and important lines must be added to the observations set: the {{Ly-$\alpha$ line of \ion{H}{1} at 121\,nm}} and the {{\ion{Mg}{2} h and k lines near 280\,nm}}. These two lines have proven their potential during rocket flights of the CLASP instrument (see Sect.\,\ref{S:method}) and are well studied theoretically. A line from the transition region into the corona, such as {{\ion{C}{4} near 155\,nm,}} would also add valuable information on the connection of the chromosphere to the corona \cite[e.g.][]{2013SoPh..288..531P}. To maximise the scientific return of the multi-line inversions, the observations must also include {{visible and IR lines}} from the chromosphere, e.g.\ of \ion{Ca}{2}, \ion{He}{1}, and visible lines from the photosphere. A gap-less coverage of the atmosphere is possible only then. This is necessary to fully characterise the magnetic field in the chromosphere.

Fluctuations of the polarization signal have to be measured with a polarimetric {{signal-to-noise ratio of 3$\times$10$^{\mathbf{3}}$}} to exploit the Hanle and magneto-optical effects in Ly-$\alpha$ and Mg~{\sc
ii}. This is based on experience with the CLASP experiment \cite[][]{2017ApJ...839L..10K} and consistent with theoretical predictions \cite[e.g.][]{2015ApJ...803...65S}.

\subsubsection{Global magnetic field in the corona\label{S:req:corona}}

The structures in the global corona are dominated by the streamer structures, coronal holes, cavities, and, if present, CMEs and their precursors. The main factor in measuring the magnetic field in the global corona is the field of view of the coronagraphic instrument. From the ground (except for rare eclipses) this is rather limited because the light that is scattered in the Earth's atmosphere limits the distance from the Sun at which coronagraphic observations can be carried out. For current ground-based coronagraphs this is less than 0.5 solar radii ($R_\odot$) above the limb \cite[COMP,][]{2008SoPh..247..411T}. Most of the large-scale coronal features extend farther out, however. A larger and more adequate field of view for a coronagraph can only be achieved by a space-based instrument. The current sheets at the top of the streamers typically start at below 1 to 2\,$R_\odot$ above the limb. The initial acceleration of a CME is also essentially achieved at these heights. The inner edge of the field of view should cover the low-lying concentrated current-carrying systems from coronal cavities and active regions. These are usually found within the first one or two pressure scale heights, the latter being about one tenth of the solar radius (for a corona at about 1\,MK). Consequently, the {{field of view  should cover 1.1 -- 3\,$R_\odot$}} from disc centre.

\begin{figure}[t]
\centerline{\includegraphics[width=0.8\columnwidth]{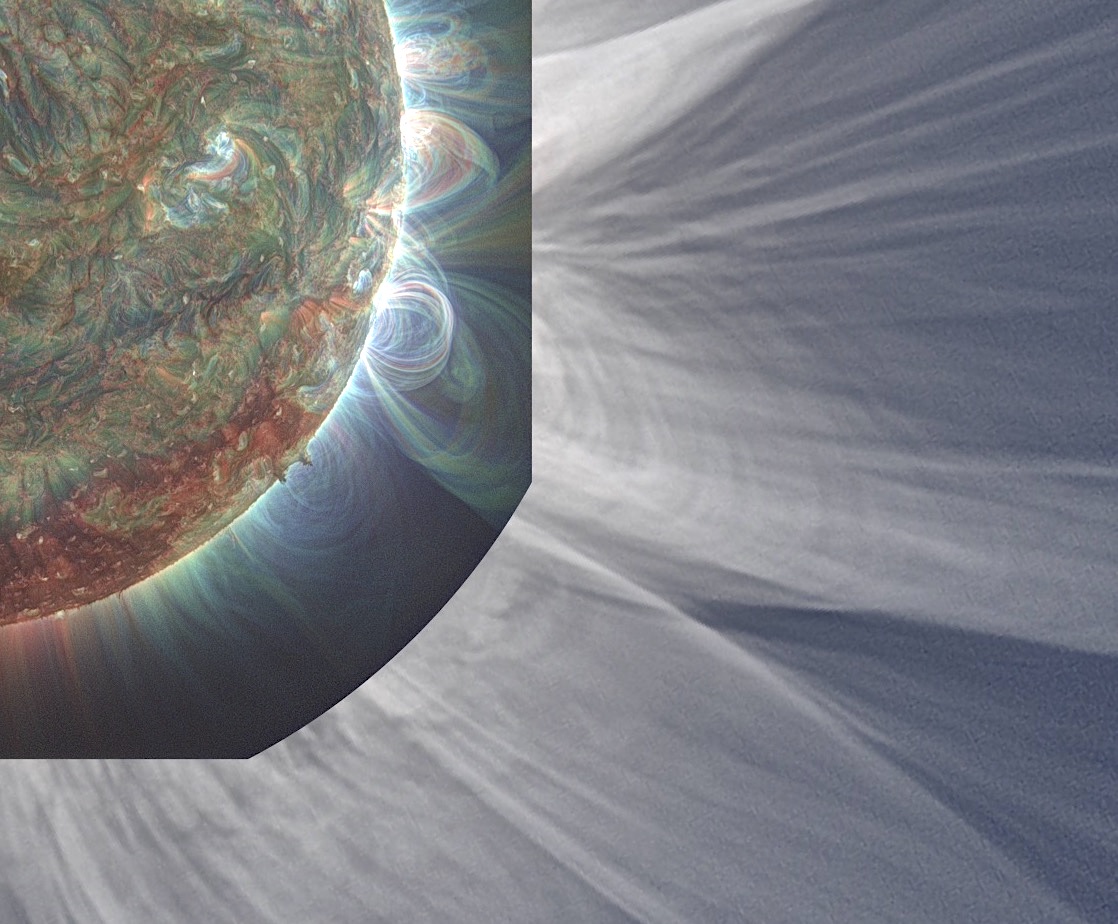}}
\caption{%
Composite image of the Sun during the eclipse on 20 March 2015. The inner part exhibits a quadrant of the solar disk showing the corona in the extreme UV by AIA/SDO (multi-colour representation of AIA channels; blue/cyan: 211\,\AA, red: 171\,\AA; green: 193\,\AA).
The outer part shows the corona as seen during totality from the ground.
The loops from the extreme UV image are seen to extend into the outer corona. Bright radial structures might represent current sheets created by magnetic field gradients.
Image credit: N. Alzate. For details on image processing see \cite{2017ApJ...848...84A}.
AIA Data courtesy of NASA/SDO and AIA science team.
Eclipse image by Druckm{\"u}ller, Habbal \& The Solar Wind Sherpas.
\label{F:eclipse}
}
\end{figure}

The requirements for spatial resolution are not as strict for the global corona as they are for the chromospheric observations. The magnetic field expands towards the outer corona, so that the spatial scales in the outer parts are much larger than near the chromosphere. In the photosphere magnetic concentrations with a diameter of a few 100 km have magnetic field strengths of more than 1\,kG \cite[][]{2010ApJ...723L.164L}. Coronal loops with lengths of 100\,Mm have field strengths of still more than 5 to 10\,G \cite[][]{2004ApJ...613L.177L}. A 100\,Mm long semi-circular loop will reach only 30\,Mm height, which is less than 0.05\,$R_\odot$. In the field of view of the coronagraph we therefore expect no more than 1\,G to a few 10\,G \cite[][]{2016NatPh..12..179J}. Conservation of magnetic flux implies that a photospheric flux tube with a diameter of 200\,km would then expand to a cross section of some 2 to 6 Mm, corresponding to up to 8\arcsec.
 A spatial resolution of about 10\arcsec\ might be sufficient to characterise the magnetic field in these elementary flux tubes, but finer threads will be seen in emission, even though the magnetic field structure is significantly smoother. This is well known for structures in the lower atmosphere and is found in models \cite[e.g.][]{2013SoPh..288..531P}. The spatial resolution to be achieved in the intensity mapping should therefore be higher. Thin threads seen in eclipse observations, for instance, show that a resolution of at least 2\arcsec\ is required.
In addition, recent imaging observations have revealed plasmoids of $\sim$2\arcsec\ that form and coalesce into larger plasmoids that expand into a CME bubble \cite[][]{2019SciA....5.7004G}. Investigating the dynamics of these plasmoids using coronagraphic observations up to 3\,$R_\odot$ will provide inestimable information on the initial acceleration of CMEs.
Combining this, the coronagraphic observations have to provide an {{angular resolution of 2 to 10\arcsec}}, depending on whether the observations are in intensity only, or if they are used to measure the state of polarisation to derive the magnetic field.

The cadence for the coronagraphic observations is governed by dynamic phenomena. A CME can reach 1000\,km/s or even more when it is accelerated from the low corona to about 1 to 2\,$R_\odot$ above the limb. Such a transient can cross the field of view of the coronagraph in about 600\,s. In order to capture the acceleration phase, the observational cadence for the magnetic field measurements should be at least 100\,s. The intensity-only observations should be acquired faster, so that we can follow the disturbances that cross the resolution element. This means that the {{temporal resolution should be in the range of 5 to 100\,s}}, again depending on the observation mode.

To derive the magnetic field in coronagraphic observations, ground-based observations have shown the high potential of IR lines such as {{\ion{Fe}{13} near 1075\,nm. \ion{He}{1} at 1083 nm}} is also sensitive to the weak coronal magnetic field \cite[][]{2010ApJ...722.1411M}. 
When the Hanle effect is used, UV lines have very high potential, in particular \Lya. So far, they have been used successfully with SOHO/UVCS \citep{1995SoPh..162..313K}, but without the polarisation capability required to derive the coronal magnetic field.
Density information crucial for interpreting the extreme-UV data, which calls for the additional capability of polarised brightness observations in the visible light, {{VL (pB)}}. 
A coronagraph should therefore cover a wide range of wavelengths from the extreme UV to the IR.

The required spectral resolution (or bandwidth) of the instruments can be estimated based on current instruments for coronagraphic observations in the UV \cite[METIS;][]{Antonucci-etal:2019} and IR \cite[COMP;][]{2008SoPh..247..411T}. These give an requirement for the bandwidth of the narrow-band imaging of {{5\,nm in the UV and of 0.1\,nm in the IR}}.
A {{signal-to-noise ratio of 100}} is required to measure the coronal magnetic field in the UV \cite[][]{1992SPIE.1546..402F, 1999SPIE.3764..147F} and IR.

\subsubsection{Connecting local and global\label{S:req:connect}}

Observations at medium resolution are required to understand the cross-scale coupling between the small-scale features of the chromosphere to the large-scale magnetic features. These observations also have to show the corona in front of the solar disc in order to connect them to the on-disc chromospheric observations. This can be achieved by imaging in the extreme UV. Because the (black-body) radiation from the surface is negligible in the extreme UV, these images will show the corona in front of the disc and above the limb.
One example would be on-disc observations that are performed near the limb. The chromospheric observations (Sect.\,\ref{S:req:chromo}) would then show the roots of the global coronal magnetic structure (Sect.\,\ref{S:req:corona}), and an extreme-UV observation would show the essential connection between the two (see Fig.\,\ref{F:eclipse}).

\begin{figure}[t]
\centerline{\includegraphics[width=0.8\columnwidth]{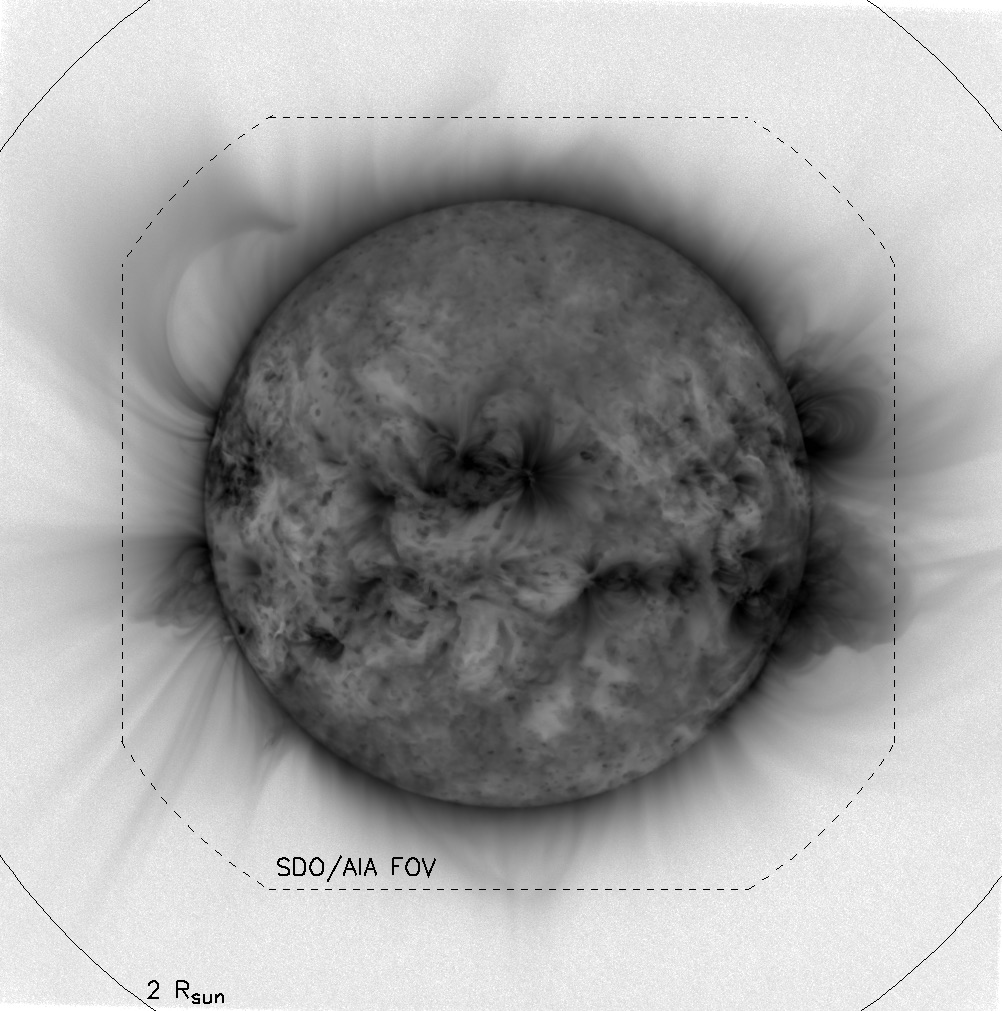}}
\caption{%
Sample image of the corona in the extreme UV as seen by SWAP/Proba\,2 (inverse colour scale). The field of view is cropped to ca.\ 0.8$^\circ${$\times$}0.8$^\circ$ (see\ Sect.\,\ref{S:req:connect}). The inner box shows the coverage of AIA, and the dotted lines in the corners mark a distance of 2\,$R_\odot$ from disc centre \cite[Image credit: ROB; see instrument description in][]{2013SoPh..286...43S}.
\label{F:SWAP}
}
\end{figure}

Coronal observations in the extreme UV are routinely performed by AIA on SDO \cite[][]{2012SoPh..275...17L}. SWAP on Proba\,2 showed that imaging in the extreme UV can also record the corona out to 2\,$R_\odot$ \cite[][]{2013SoPh..286...43S}. To ensure sufficient overlap with the coronagraphic observations (Sect.\,\ref{S:req:corona}), the imaging in the extreme UV should cover the corona across the whole disc out to 1.5\,$R_\odot$ (see\ Fig.\,\ref{F:SWAP}). This corresponds to a {{field of view of 0.8$^\circ${$\times$}0.8$^\circ$}}.

Coronal features seen in the extreme UV have a distribution of widths, with a peak around 550\,km and a tail to broader features \cite[][]{2017ApJ...840....4A}. A {{spatial resolution of 1\arcsec}} should therefore be achieved; this is slightly better than that of AIA. 

With the same arguments as above, the cadence of the observations should reach 3\,s to follow changes in intensity. Polarimetric observations of the magnetic field can have a lower cadence of 60\,s because the magnetic field evolves slower than the thermal properties of the corona. The {{temporal resolution therefore must be in the range of 3 to 60\,s}}.

A very good choice for the observations is the {{coronal line of \ion{Fe}{10} at 17.4\,nm}}. To isolate the line, the {{width of the band has to be 0.35\,nm}}. The emission from this ion shows high contrast in coronal structures, as is well know from AIA and SWAP. Because the line lies in the saturated Hanle regime,  it will allow determining the direction of the magnetic field through an analysis of the linear polarisation. Based on an earlier study, the required {{signal-to-noise ration is $10^3$}} \cite[][]{2012ExA....33..271P}.

\subsection{Instrumentation required to perform measurements of the magnetic
field in the upper atmosphere\label{S:instruments}}  

The two quite distinct measurements requirements are to record the small-scale magnetic fields that are observed in the chromosphere and the large-scale magnetic field in the solar corona. They necessitate quite different techniques: the former needs observations on the disc at high resolution, and the latter coronagraphic observations by occulting the solar disc. These tasks require two separate instruments. A third instrument is needed to connect the measurements of these first two instruments.  More details and a possible mission scenario are discussed in Sect.\,\ref{S:mission}.

{\bfseries{1)~  UV-to-IR telescope with a large aperture.}}~
A UV-to-IR telescope with an aperture of about 3\,m is required to achieve the necessary signal-to-noise ratios to derive the magnetic field in the solar chromosphere from the polarisation measurements. The required spectral lines in the UV can be recorded only in space, and only the stable space environment will allow the high-performance observations required in the IR. Equally important, the high spatial resolution over a large field of view is only possible from a space-based platform that is not affected by the Earth's atmosphere.

{\bfseries{2)~ Extreme UV-to-IR coronagraph.}}~
A coronagraph is required to study the large-scale magnetic field out to 3\,$R_\odot$ from disc centre. The large field of view can only be achieved from space, even for observations in the IR. UV observations in Ly-$\alpha$ provide high sensitivity to the magnetic field by use of the Hanle effect. An aperture of 40\,cm is required to achieve the necessary signal-to-noise ratio.

{\bfseries{3)~ Extreme UV imaging polarimeter.}}~
An instrument is required to observe the corona in front of the disc and above the limb to connect the observations of the other two instruments. This requires an aperture of about 30\,cm to achieve the signal-to-noise ratio that is necessary to also derive the magnetic field direction from polarimetric observations.

\section{Mission scenario for measuring magnetic fields throughout the  solar atmosphere\label{S:mission}}

  One notable constraint for a possible mission  scenario is the size of the telescope for chromospheric observations. Our considerations in Sects.\,\ref{sec:obs_chromosphere} and \ref{sec.m.class} will show that this should have an aperture of 1 to 3\,m, which means that this instrument would dominate the configuration of the space observatory. Because of its size, it will have to be part of the spacecraft and thus be under ESA responsibility. Post-focus instrumentation such as a spectropolarimeter and an imager can then be considered as payload provided by institutions throughout ESA member states.
  The other two telescopes, the coronagraph and the EUV polarimetric imager (Sects.\,\ref{sec:obs_corona_offlimb} and \ref{sec:obs_corona_ondisc}), can be mounted on the sides of the main telescope. This ensures that the coronagraph aperture has an unobstructed view over 2$\pi$ steradians to fulfil
the stringent stray-light requirements of this instrument.

  We assume that a near-Earth orbit would be adequate for such a mission, even though a position at L1 would provide a continuous coverage of the Sun and a very stable environment. Observations from L4 or L5  would provide the highly interesting possibility of conducting coordinated observations with Earth-based facilities: viewing the Sun at the same time from two different angles would enable stereoscopic observations. Determining the orbit that is best suited for the mission will require a careful trade-off study based on the available technology in data transmission, spacecraft pointing stability, thermal environment control, and science goals.

   In Sects.\,\ref{sec:obs_chromosphere} to \ref{sec:obs_corona_ondisc} we provide  details of straw-man instruments (cf.\ \sect{S:instruments}). In Sect.\,\ref{sec.m.class} we discuss options for a possible M-class mission scenario.

\begin{figure*}[t]
\centerline{\includegraphics[width=\textwidth]{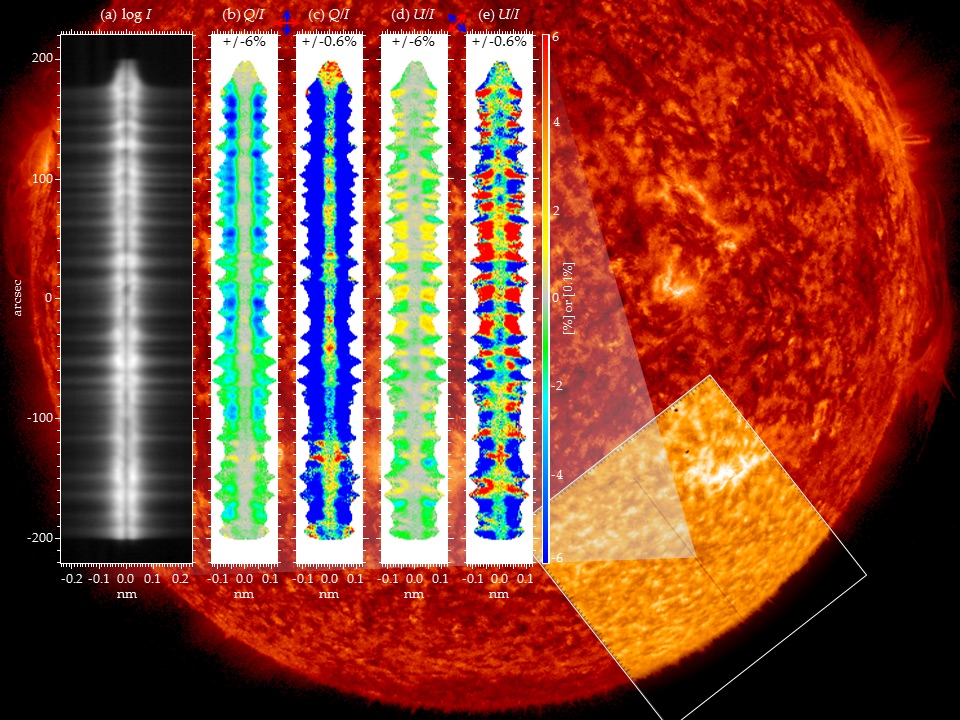}}
\caption{%
Spectropolarimetric measurements with CLASP from its first sub-orbital flight on 3 Sept.\ 2015. The background image shows the chromosphere of the Sun, and the inset displays the slit-jaw filtergram acquired by CLASP in the Ly-$\alpha$ line at 121\,nm. Panels (a) to (e) show the spectrum along the slit in Ly-$\alpha$ in total intensity (left) and in different orientations of the linear polarization  
\cite[image credit NAOJ, JAXA, NASA/MSFC, IAC, IAS; taken from https://www.nao.ac.jp/en/news/science/2017/20170518-clasp.html; based on][]{2017ApJ...839L..10K}.
\label{F:clasp}
}
\end{figure*}

  \subsection{Measuring the magnetic field in the chromosphere}
  \label{sec:obs_chromosphere}

  The primary requirement for spectropolarimetric observations is the signal-to-noise ratio at the spatial and temporal scales to be resolved.  Following the arguments presented in the previous sections, true advances in the study of the magnetic field throughout the solar chromosphere will be obtained with spectropolarimetric measurements in the key lines Mg~h \& k near 280\,nm and \Lya\ at 121\,nm at spatial scales of the order of 0.1\arcsec\ (about 70 km) and temporal scales of the order 1 to10~s.  Accurate measurements of the magnetic field at these scales will be obtained with a signal-to-noise
ratio in the Stokes parameters of the order of $2\times 10^3$.  Finally, another important requirement is the ability to map the magnetic field structure over areas that are comparable to typical large-scale magnetic structures, e.g.\ active regions (about 300$\times$300\arcsec). A spatial resolution of  0.1\arcsec\ requires a spatial sampling of 0.05\arcsec\ per pixel. Thus a field of view of 300$\times$300\arcsec\ requires a 6k$\times$6k sensor. This is only marginally larger than the 4k$\times$4k sensors that are used by SDO and are expected to be easily available in the next decade.

Our estimates for the chromospheric telescope are based on the performances of current and near-future facilities that are designed to measure the solar magnetic field at high resolution. The suborbital rocket flights of CLASP   \citep{2017ApJ...839L..10K} in particular can be regarded as the true pathfinders for this mission
(see \fig{F:clasp}). CLASP included spectropolarimetric and imaging capabilities (i.e. slit-jaw imagers) to analyse \Lya\ and \ion{Mg}{2}.
  The two CLASP flights demonstrated that the required polarimetric accuracy at the 1\arcsec\ spatial and 1~s temporal resolution can be reached with an aperture of about 30~cm. When we scale this to the  0.1\arcsec\ spatial resolution required here, this calls for an effective area that is 100 times larger. When we assume no (significant) increase in throughput for the whole instrument, this would require a 3~m class telescope.
In the IR, e.g. for observing \ion{He}{1} at 1083\,nm, the diffraction limit of such an instrument is about the required spatial resolution of 0.1\arcsec. At the wavelength of \Lya,\ the required resolution of 0.1\arcsec\,is  less stringent  than the theoretical diffraction limit, so that the design of the telescope will be challenging but possible within the envelope of an L-class mission.  Moreover, the required resolution of 0.1\arcsec\ can be maintained over the required large field of view.

  The science requirements for an instrument like this demand spectropolarimetric observations at the \Lya\ and the Mg~II h \& k lines, but also in the visible and IR (see \tab{tab:sci.req}). Currently available Al-based coatings with high reflectivity in the visible (also required for the thermal control of the instrument) easily allow  observing lines from the IR down to the UV at about 110~nm. At the required resolving power of $\sim 3\times10^4$ (over two pixels) at 120~nm, and with a 6k$\times$6k detector, the \Lya\ line can be observed simultaneously with several very important lines such as the Si~III and 120.6~nm and the N~V lines at 123.8 and 124 nm. 
  The resolving power at the Mg~II h \& k doublet, on the other hand, would be of the order of $6\times10^4$.

A separate post-focus instrument would be required to cover the visible and IR lines to be observed. Here, well-known concepts from existing and upcoming ground-based facilities can be used to reach the required specification. We do not discuss these further.

  \subsection{Measuring the magnetic field in the off-limb corona}
  \label{sec:obs_corona_offlimb}

  The instrument for on-disc spectropolarimetry described above should be paired with a coronagraph capable of observing
the large-scale corona.
 Again, the \Lya\ line should be observed together with IR lines (see \tab{tab:sci.req}), now in narrow-band polarimetric imaging. Broad-band measurements of visible-light polarised brightness, called pB, are also essential for the coronagraph.  They will provide the coronal electron density, a basic ingredient for an analysis based on which the coronagraphic spectropolarimetric observations are interpreted.  

  The primary requirements on the instrument design are again the signal-to-noise
ratio in polarimetric measurements and the field of view. The latter should have an inner edge as low as 1.1 $R_\odot$ to have a significant overlap with the on-disc polarimetric imager (\sect{sec:obs_corona_ondisc}) and an outer edge out to 3 $R_\odot$ to capture the large-scale corona   that is not visible from the ground. The spatial, temporal, and spectral resolution requirements are less stringent than those for the chromospheric telescope (see Table~1).
  In terms of coronagraphic performances, any residual stray light should be lower than the disc brightness by a factor $10^{-9}$ in the visible and IR and $10^{-7}$ in \Lya.

  The baseline for such an instrument would be a dual-band coronagraph capable of obtaining large field-of-view narrow-band polarimetric images in the UV (\Lya) and in the visible and IR ranges. The set of visible and IR lines to be considered for would include \ion{Fe}{14} at 530.3\,nm (``green line''), \ion{Fe}{10} at 637.6\,nm (``red line''), \ion{Fe}{11} at 789.2\,nm, and the \ion{Fe}{13} doublet and \ion{He}{1} lines around 1080 nm, for instance.

  An instrument such as this has a strong archetype in the Metis coronagraph \citep{Antonucci-etal:2019}, which is now flying on Solar Orbiter. With respect to Metis, the proposed instrument will have polarimetric capabilities in \Lya\ and in a selection of visible and IR lines, a field of view extending down to 1.1 solar radii instead of 1.6, and higher spatial resolution (especially at \Lya). Scaling the Metis instrument to the requirements outlined here in \tab{tab:sci.req} results in a coronagraph with an aperture of about 40\,cm.

  \subsection{Measuring the magnetic field in the on-disc corona}
  \label{sec:obs_corona_ondisc}

This instrument is mainly required to characterise the magnetic field in the million Kelvin corona and to provide context and connection information for the chromospheric  telescope (\sect{sec:obs_chromosphere}) and the coronagraph (\sect{sec:obs_corona_offlimb}). This is essentially an improved  development of imagers that were used for SDO/AIA or Proba\,2/SWAP, for example, but now with the capability of recording the polarisation signal.

The field of view  and resolution requirements (\tab{tab:sci.req}) call for a 6k$\times$6k detector and a focal length of 4.4 m with 12 $\mu$m pixels. With currently available 4k$\times$4k detectors, we can expect the required detectors to become available in the coming years. A single on-axis parabolic mirror would meet the resolution requirement, but over a limited field of view, and would be overly bulky. Even though it adds one reflection, the preferred option is therefore a Ritchey-Chretien telescope. The linear polariser is a 45$^\circ$ folding mirror that is located before the detector.

The 45$^\circ$ polariser is located close to the detector (about 10\,mm), and the whole assembly is rotated to provide the necessary  measurements of linear polarisation. A single-mirror polarising system is required to maximise the photon count in a given exposure time. This is achieved by a rotating focal plane assembly, which implies that any connectors follow the movements of the detector.
A two-mirror polariser would not require a rotating focal plane assembly, but would  reduce the flux at the focal plane by a factor two.

To select the main target line of \ion{Fe}{10} at 17.4\,nm, multi-layer coatings can be optimised for maximum throughput at 17.4 nm and maximum rejection of other nearby emission lines, some of which are predicted to show zero or opposite polarisation compared to \ion{Fe}{10}. It is important to reject these lines because they constitute stray light in the polarisation measurement and degrade the measurement signal-to-noise
ratio. Such coatings have already been developed for EUI/Solar Orbiter. Using new techniques for the multi-layer mirror coatings, we can reach a width of the transmitted band of only 0.35\,nm, which is twice as narrow as bands on current instruments.

\subsection{Mission scenario in case of an M-class mission opportunity\label{sec.m.class}}

What we outlined so far in this paper calls for an L-class mission. We consider that the chromospheric telescope with an aperture of up to 3\,m is the main cost driver in our mission profile. When we downsize our mission scenario for an M-class mission, the 3\,m telescope is therefore the first item to consider. Other mission options can also open up when a smaller main telescope is considered.

The effect of reducing the size of the chromospheric telescope from a 3\,m class to a 1\,m class instrument means cuts in the science return. In addition to a reduced spatial resolution, the main downgrade would be the sensitivity to the magnetic field, and here in particular in the UV, where the aperture is essential for reaching the required signal-to-noise ratio. The smaller telescope will reduce the spatial resolution by more than what is suggested by the reduction of the aperture (to collect the same number of photons per resolution element), or the sensitivity to the magnetic field will be reduced, or both. A trade-off study will have to show possible paths when the aperture needs to be reduced and what effects these would have for the science goals.

Even a 1\,m class telescope would outperform any currently operational or planned solar space observatory. We can still expect significant advances. The scientific return would be downgraded from a full characterisation of the magnetic field in the outer atmosphere, however. We would mostly be able to investigate only bright structures on the Sun, while the faint features are often essential to understanding the global large-scale connection. This would affect all of the science goals outlined in \sect{S:science} because the magnetic field is present not only in the bright features, but is volume filling. It would mean that we could only investigate a fraction of the outer solar atmosphere. 

Downgrading the chromospheric telescope from a 3\,m to a 1\,m class instrument might also open new opportunities. For example, one could place a smaller observatory at either L4 or L5 instead of in Earth orbit (or at L1). In combination with a facility either in Earth orbit or on the ground, this  would give the possibility for stereoscopic observations.

\section{Technological challenges}

\subsection{Main telescope: Thermal control and optical performance
over the field of view}

Relatively large telescopes that cover the visible to UV range have previously been
flown into space or near-space (stratospheric) environments, for example the 0.5\,m telescope on the Hinode telescope \cite[][]{2008SoPh..249..197S} 
or the Sunrise balloon-borne 1\,m telescope 
\cite[][]{2011SoPh..268....1B}.
These two instruments have proven that it is possible to build a UV-to-visible
light solar telescope that manages the solar flux while keeping performances
up to the tight requirements (e.g. the Sunrise telescope is diffraction
limited in the visible). 
In the case of the 3 m telescope discussed here, a Gregory design with
a heat rejection field stop will also have to be used. The telescope can be seen
as a scaled-up version of the telescope on board Sunrise. Because of the size, it will be
challenging but possible, especially when we consider that it does not have to be diffraction
limited.

The requirement of maintaining the image quality over
the large (300\arcsec$\times$300\arcsec) field of view needs some consideration, however.
A simple Gregory design with two powered mirrors cannot maintain the required
0.1\arcsec\ resolution (two pixels) over the full field of view. A corrective optical
element or a curved focal plane assembly will have to be considered.

\subsection{Data transmission, handling, and storage}

\begin{figure}
{\includegraphics[width=\textwidth]{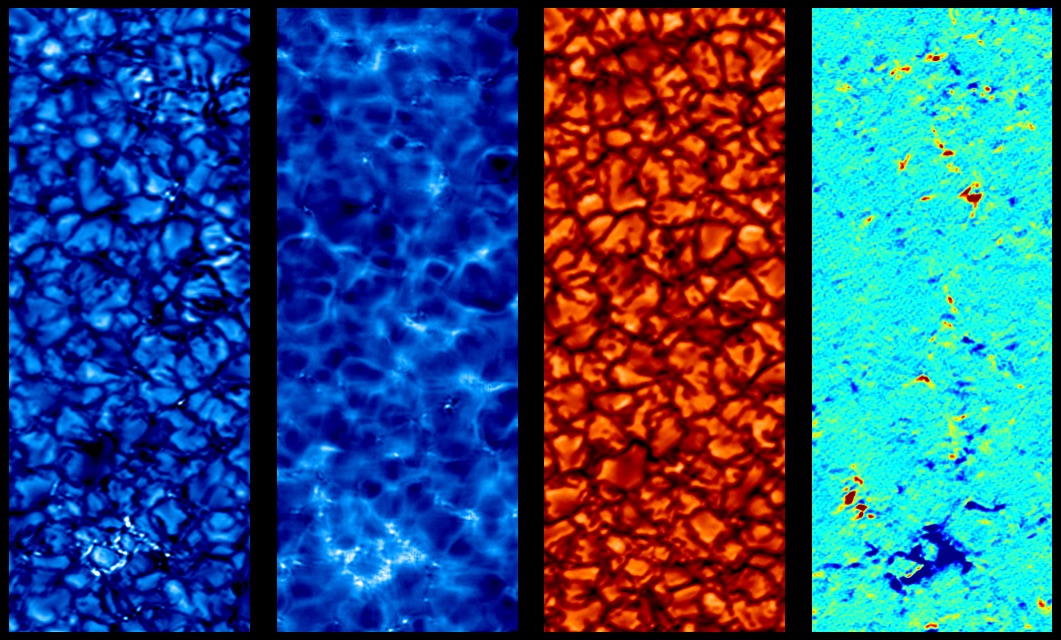}}
\caption{%
Multi-wavelength observation of the solar photosphere and chromosphere. From left to right the maps recorded by the Sunrise balloon telescope show: UV at 300\,nm, \ion{Ca}{2} K at 397\,nm, \ion{Fe}{1} at 525\,nm and the photospheric magnetic field (red and blue are opposite magnetic polarities). The field of view is very small, ca.\ 13\arcsec$\times$38\arcsec. We propose to acquire data in multiple lines in a much larger field of view with a continuous high data rate. This would produce more than 50\,TB within 8 hours. Image based on \cite{2010ApJ...723L.127S}.
\label{F:data}
}
\end{figure}

The   number of large-format cameras will be significant: at least three for
the chromospheric telescope, two for the coronagraph, and one for the
 EUV polarimetric imager. They have to operate simultaneously with a cadence of down to 1\,s, so that even with some loss-less data compression, this leads to a data
rate of the order of 1 Gbit/s (see \fig{F:data}).
A data rate like this is certainly challenging and will require the use of 
high-capacity transmission systems such as the laser communication systems
currently under development and testing. It is likely that such systems are available
in one or two decades.

Large on-board storage units are required that have a capacity to temporarily save 50 TB or more. This is the equivalent of continuously observing for 8 hours.  Such a storage unit requires an improvement in capacity by a factor of 10 compared to\ PHI/Solar Orbiter, for instance, which might be feasible in a decade from now. 

The continuous stream of data at a high rate requires re-thinking the techniques for processing and archiving data. While the storage of this amount of data might not pose a significant challenge in a decade, the processing will need significant improvements. Here machine-learning techniques to be developed in the framework of big-data strategies  
 will help with the efficient and automated standard processing steps, and may even provide a (preliminary) data selection.

\section{Relation to other space- and ground-based observations\label{S:others}}

Synoptic observations of the whole solar disc and the corona (at moderate resolution) are important to place high-resolution data acquisition as proposed in this paper into context.
Synoptic observations are currently available through a range of space- and ground-based facilities, for example by SOHO/LASCO, SDO, Proba2, STEREO, Hinode, or GONG+. Because of the relevance for operational space weather services, it can be assumed that (similar) global monitoring observations will remain continuously present in the future (GOES, SWFO, ESA Lagrange). This will guarantee that the high-resolution data we propose in this paper for the time frame of 2035 to 2050 will be supplemented by proper context information.

The near future will see a significant shift in ground-based solar telescopes from the 1\,m class to the 4\,m class.
Current high-resolution ground-based solar observations (GREGOR, SST, GST, NVST, ONSET) are complemented by unique high-resolution data from occasional balloon flights (Sunrise), all with apertures from 1\,m to 1.5\,m.
The Daniel K. Inoue Solar Telescope (DKIST) with 4\,m aperture has just seen first light. 
DKIST observes the solar disc, and has a coronagraphic mode for the innermost part of the corona. Using lines such as \ion{Ca}{2} near 394\,nm and 854\,nm and \ion{He}{1} near 1083\,nm, it will explore the magnetic field in the low chromosphere through on-disc measurements. Observations\ in \ion{Fe}{13} at 1075\,nm above the limb will give access to the magnetic field in the very low corona, at only a few 100\arcsec\ (less than 0.5\,$R_\odot$) above the limb.
A telescope with comparable specifications is proposed by the European solar community (European Solar Telescope, EST), for which first light is scheduled for the mid- to end-2020s.

The 3\,m class telescope we propose here for a space-based platform will provide a great leap in scientific exploitation compared to the upcoming 4\,m class ground-based telescopes.
The wavelength range can be extended farther into the UV, which is not accessible from the ground. This is indispensable to obtain access to lines such as Ly-$\alpha$ at 121\,nm or \ion{Mg}{2} near 280\,nm and thus measure the magnetic field in the uppermost parts of the chromosphere. A complete coverage and a holistic view of the whole outer solar atmosphere becomes available only then.
The high-resolution field of view of ground-based observations will always be limited by seeing effects that are introduced by the Earth's atmosphere. With adaptive optics systems, the corrected high-resolution region is roughly 10\arcsec\ across, multi-conjugate adoptive optics,  might extend this to some 50\arcsec. Only a seeing-free space-based platform can provide the required large field of view of 300\arcsec\ across at constant high resolution, however.
Furthermore, only a space platform can provide stable observing conditions to follow structures over hours and to minimise the cross-talk between the different polarisation states.

Coronagraphic measurements from the ground are marred by the inevitable stray light in the Earth's atmosphere. Among other effects, this limits the coverage of the field of view to a maximum of 0.5 solar radii (optimistic estimate) above the limb for DKIST. While DKIST can observe the coronal magnetic field at a high spatial resolution of about 0.1\arcsec, it will not be able to investigate the large-scale magnetic field. To reach significantly farther out, we have to enter space.

The next important step for the solar community in space is provided by the Solar Orbiter mission that was launched in February 2020. This mission will address many of the fundamental questions in solar physics, in particular those concerning the connection from the Sun to the heliosphere. However, it will not directly measure the magnetic field in the outer solar atmosphere. The science questions outlined in this paper in \sect{S:science} will remain largely open after Solar Orbiter. EUVST, a Japan-led high-resolution extreme-UV spectrometer, is planned to be launched in the late 2020s. This will also lack the capability of investigating polarisation and thus the magnetic field in the outer solar atmosphere.

In this paper we suggest to measure the magnetic field  to determine what drives the
dynamics in the outer solar atmosphere. The full science return can only be achieved by an L-class mission. With a downgrade to an M-class mission (see \sect{sec.m.class}), other  scenarios with a broader scope might become possible.
One option that could be instigated would be an L-class mission in the context of Voyage2050 that would combine our suggestion (downgraded to M-class) with one or more missions in the fields of solar, heliospheric, magnetospheric, and ionospheric physics as described in other papers in this volume. This would provide a Grand European Heliospheric Observatory and ensure scientific advances in the coming decades in a broad scientific context.

\section{Conclusions} 

The Sun is the star closest to Earth. It is the only star for which we can resolve
details on its surface and in its atmosphere. The outer solar atmosphere from
the cool chromosphere to the hot corona is permeated by a magnetic field that
energetically dominates the gas. The magnetic
field controls the level of solar activity. It thus modulates the radiative
output of the Sun throughout the activity cycle, largely governs the emission
from the UV to X-ray radiation,  ejects particles and fields  from the
corona, and drives space weather effects. Through  these magnetic
activity effects, the Sun directly impacts the Earth and human life. Likewise,
the magnetic activity of other stars will have an impact on the planets orbiting
these host stars, and it will govern the habitability on these planets.

Although we can observationally characterise the magnetic field on the solar
surface (the photosphere) reasonably well, we fail to do so for
the higher layers. In the overlying chromosphere  and in the corona that  stretches
essentially into interplanetary space, our information
on the magnetic field is mainly indirect. This state
of knowledge is highly unsatisfactory because the magnetic field energetically dominates the plasma in this outer solar atmosphere and thus largely
controls the phenomena that are observed there. Only by knowing the magnetic field and
its evolution can we unlock the source  of heating of the gas to millions
of degrees, understand the shape
and brightness of the corona, learn about magnetic reconnection and its effects such as flares and the associated energetic particles and radiation,
and unravel the initiation of large-scale
ejections of parts of the solar atmosphere.

The lack of direct measurements of
the magnetic field in the outer solar atmosphere is a major hindrance to our
understanding of the physics of the outer solar atmosphere. If we could remedy this, we would be able to
explain a very wide variety of phenomena.
In essence, we are blind to the main driver
of solar activity in the critical layers of the solar atmosphere. 

By directly measuring the magnetic field in the outer atmosphere of the Sun,
we will be able to address four paramount long-standing questions of solar physics:

\begin{itemize}

\item[1.] How does the magnetic field couple the different layers of
the atmosphere, and how does it transport energy?

\item[2.] How does the magnetic field structure, drive and interact with the plasma in the chromosphere and upper
atmosphere?

\item[3.] How does the magnetic field destabilise the outer solar atmosphere
and thus affect the interplanetary environment?

\item[4.] How do magnetic processes accelerate particles to high energies?

\end{itemize}

To address these science questions, we have to perform new ground-breaking
observations with a suite of space-based solar telescopes that far exceed
current capabilities in terms of spatial resolution, light-gathering power,
and polarimetric performance. This requires instruments for different parts
of the outer solar atmosphere with sufficiently large apertures.

\begin{itemize}

\item[a.] Large-aperture UV-to-IR telescope.
Depending on the mission scenario (M or L), this telescope of the 1-3 m class is aimed mainly to measure the magnetic field in the
chromosphere by combining high spatial resolution and high sensitivity.

\item[b.] Extreme-UV-to-IR coronagraph.
A coronagraph that is designed to measure the large-scale magnetic field in the corona. The required aperture is about 40\,cm.

\item[c.] Extreme-UV imaging polarimeter.
A 30\,cm telescope that combines high throughput in the extreme UV with polarimetry
to connect the magnetic measurements of the other two instruments.
\end{itemize}

All three instruments will perform novel measurements of the magnetic
field in the outer atmosphere that currently cannot be achieved.
This solar  observatory could be placed in a near-Earth orbit to maximise
the data downlink, or at L4 or L5 to provide stereoscopic observations of
the Sun in combination with Earth-based observatories.

This mission to measure the magnetic field will unlock the driver of the dynamics in the outer solar atmosphere. Through this it will greatly advance our understanding of the Sun and the heliosphere.


\begin{acknowledgements}
E.A.B. and L.B. gratefully acknowledge financial support from the Swiss National Science Foundation (SNSF) through Grant 200021\_175997
S.P. acknowledges  the  funding  by  CNES  through  the MEDOC  data  and  operations  center, and the funding by ESA through the Virtual Space Weather Modelling Centre - Part III. 
J.T.B. acknowledges the funding received from the European Research Council (ERC) under the European Union?s Horizon 2020 research and innovative programme (Advanced Grant agreement No. 742265).
\end{acknowledgements}


\bibliographystyle{spmpsci}

\bibliography{voy2050}

\end{document}